\documentclass[10pt,twocolumn,english,aps,manuscript,prb,floatfix,superscriptaddress,twocolumn,showpacs,amsmath,amssymb,reprint]{revtex4-1}
\usepackage[T1]{fontenc}
\usepackage[latin9]{inputenc}
\usepackage{color}
\usepackage{amsmath}
\usepackage{amssymb}
\usepackage{graphicx}
\usepackage{esint}
\makeatletter

\providecommand{\tabularnewline}{\\}

\@ifundefined{textcolor}{}
{%
 \definecolor{BLACK}{gray}{0}
 \definecolor{WHITE}{gray}{1}
 \definecolor{RED}{rgb}{1,0,0}
 \definecolor{GREEN}{rgb}{0,1,0}
 \definecolor{BLUE}{rgb}{0,0,1}
 \definecolor{CYAN}{cmyk}{1,0,0,0}
 \definecolor{MAGENTA}{cmyk}{0,1,0,0}
 \definecolor{YELLOW}{cmyk}{0,0,1,0}
}

\PassOptionsToPackage{caption=false}{subfig}

\@ifundefined{showcaptionsetup}{}{%
 \PassOptionsToPackage{caption=false}{subfig}}
\usepackage{subfig}
\makeatother

\usepackage{babel}
\begin{document}

\title{Competing Adiabatic Thouless Pumps in Enlarged Parameter Spaces}

\author{Pedro L. e S. Lopes}

\email{plslopes@ifi.unicamp.br}

\affiliation{Instituto de F\'{i}sica Gleb Wataghin, Universidade Estadual de Campinas,
Campinas, SP 13083-970, Brazil}

\author{Pouyan Ghaemi}

\affiliation{Physics Department, City College of the City University of New York, New York, NY 10031, USA}

\author{Shinsei Ryu}

\affiliation{Department of Physics and Institute for Condensed Matter Theory,
University of Illinois, 1110 W. Green St., Urbana IL 61801-3080, USA}

\author{Taylor L. Hughes}

\affiliation{Department of Physics and Institute for Condensed Matter Theory,
University of Illinois, 1110 W. Green St., Urbana IL 61801-3080, USA}

\begin{abstract}
The transfer of conserved charges through insulating matter via smooth deformations of the Hamiltonian is known as quantum adiabatic, or Thouless, pumping. Central to this phenomenon are Hamiltonians
whose insulating gap is controlled by a multi-dimensional (usually two-dimensional) parameter space in which paths can be defined for adiabatic changes in the Hamiltonian, i.e., without closing the gap. Here, we extend
the concept of Thouless pumps of band insulators by considering a larger, three-dimensional parameter space. We show that the connectivity of this parameter space is crucial for defining quantum 
pumps, demonstrating that, as opposed to the conventional two-dimensional case, pumped quantities depend not only on the initial and final points of Hamiltonian evolution but also on the class of the chosen path and
preserved symmetries. As such, we distinguish the scenarios of closed/open paths of Hamiltonian evolution, finding that different closed cycles can lead to the pumping of different quantum numbers, 
and that different open paths may point to distinct scenarios for surface physics. As explicit examples, we consider models similar to simple models used to describe topological insulators, but with doubled degrees of freedom compared to a minimal topological insulator model.  The extra fermionic flavors from doubling allow for extra gapping terms/adiabatic parameters - besides the usual topological mass which preserves the topology-protecting discrete symmetries - generating an enlarged adiabatic parameter-space. We consider cases 
in one and three \emph{spatial} dimensions, and our results in three dimensions may be realized in the context of crystalline topological insulators, as we briefly discuss.
\end{abstract}
\maketitle

\section{Introduction}
Insulating matter is often thought of as a transport-inert environment. It was thus quite surprising that charge currents could be established in an insulator at zero temperature and  the concept of `adiabatic pumping', as described by Thouless, now pervades the literature\cite{ThoulessPump,NiuRMP,RestaRMP}. Such pumps characterize whether a periodic variation of a set of Hamiltonian \emph{parameters} of a gapped system leads to the transport of (quantized) charge, that is conserved due to the presence of a symmetry.

To illustrate, the conventional example considers a 1D insulator (with lattice constant $a$) under a slow, adiabatic translation - parametrized by a phase shift $\delta\phi (t)$ - of the underlying periodic potential: $V(x)\to V(x+\delta\phi(t) a/2\pi)$\cite{ThoulessPump}. During one cycle, $\delta \phi(t):0\rightarrow 2\pi$, an integer amount of electron charge is pumped through the 1D system. This integer is equal to the 2D Chern number of the 1D Bloch Hamiltonian parameterized by the 1D momentum $k$ and $\delta\phi.$ Other examples consist of spin pumps, defined by the spin Chern number when some component of the electron spin is conserved, and discrete $Z_2$ pumps (e.g., Kramers' pair, fermion parity) defined by a bulk $Z_2$ invariant\cite{FuKaneZ2,TeoKaneDefects,ParityPump}. Higher dimensional generalizations of the charge pump are also possible, the first of which being the magneto-electric polarizability pump, fixed by the second Chern number\cite{QHZ2008,Essin2009,vanderbiltnew}.

The ubiquitous relation between quantum pumping and bulk Chern numbers connects the notion of adiabatic transport in gapped systems to the concept of topological band insulators. Such topological phases constitute a class of  gapped systems which generically display gapless surface states, when in the presence of certain symmetries such as time-reversal. These surface states are inherently robust to disorder, persisting even in the presence of interactions as long as the relevant symmetries are not broken (explicitly or spontaneously) and no surface topological order is developed. This class of systems displays a short-ranged profile of spatial entanglement, and constitutes what are now known as 'symmetry protected topological phases' (SPTs)\cite{WenBook,FradkinBook}.

In addition to adiabatic pumping during \emph{closed cycles}, topological responses (electromagnetic, thermal, etc.) of SPTs may be derived by considering adiabatic transformations of the gapped Hamiltonian on special \emph{open paths} in  parameter space. These paths consist of adiabatic (gapped) interpolations between trivial and topological symmetry-preserving reference Hamiltonians\cite{QHZ2008, HosurPRB}. In order to adiabatically connect the two end points, somewhere (or everywhere) in-between, the protecting symmetry must be broken. For example, for a 1D insulator with inversion symmetry, a trivial phase has a vanishing charge polarization, modulo an integer charge. Deforming the corresponding Hamiltonian parameters into the topological regime changes the polarization to its topological value of $e/2$ (again modulo integer charge). Comparing the initial and final states, the difference is a half-integer polarization. A symmetry preserving interpolation between the Hamiltonians of an inversion symmetric system in different topological phases generically implies
the closure of the system gap somewhere during the interpolation. This gapless region is singular in the sense that it renders the notion of `adiabatic transformation' ill defined. An adiabatic/gapped interpolation between the two reference points, therefore, demands the introduction of parameters which break the protecting symmetry of the Hamiltonian (inversion in the present example). If the interpolation is continued onward to cycle back to the initial trivial state this forms a closed path and the polarization in the final state can only differ from the polarization of the initial state by an integer number of electrons; hence, indicating a quantized number of electrons pumped.

From the previous discussion, the adjective `adiabatic' should be understood in the present context as implying that for all allowed values of the parameters defining the Hamiltonian, the gap should remain open.
The description of the interesting pumping path discussed above demands a minimal set of \emph{two parameters}, the original symmetry preserving gap term that switches between the topological phases, and a symmetry breaking gap term. 
Interpolations define one-dimensional trajectories in this non-simply connected two-dimensional space (the point with both gap terms vanishing must be removed as the system is gapless there). 

In this article we focus on generalizing the notion of adiabatic quantum transport and responses to \emph{enlarged nodal parameter spaces}.  We consider, concretely, parameter spaces in one higher dimension (i.e., three parameters instead of two.)
that are non-simply connected, 
and allow for different classes of trajectories which cannot be continuously deformed into one another. The specific case we consider will be a three-dimensional parameter space with several nodal/gapless lines.
Interestingly, even when paths connect the same two reference Hamiltonians,  the different possible classes can lead to different response properties, and ultimately the pumping of different conserved charges.
This should be contrasted with the aforementioned cases -
the difference in the electromagnetic response (e.g., difference in polarization) defined for an open path with two symmetry-preserving end-points 
depends only on the end-points themselves, but not on the choice of path (modulo integer numbers of electrons).
Accordingly, all adiabatic cycles of a given Hamiltonian would pump the same quantum numbers.
This result depends on   the mathematical form of the topological invariant and the topology of the pumping parameter space itself. Here we will show that changing the topology of the parameter space can lead to interesting new features.
% the response difference induced by an open path with two symmetry-preserving end-points is shown only to depend on the end-points
%themselves and not on the choice of path.

%In most of the aforementioned articles - because of the mathematical form of the topological invariant and the topology of the pumping parameter space itself - the response difference induced by an open path with two symmetry-preserving end-points is shown only to depend on the end-points
%themselves and not on the choice of path.
%Accordingly, all adiabatic cycles of a given Hamiltonian would pump the same quantum numbers.
%As will be discussed, the parameter spaces we consider will also be non-simply connected, allowing for different classes of linear trajectories which cannot be deformed into one another. These, despite connecting the same reference Hamiltonians, will be seen to lead to pumping of different conserved charges.

While our results will apply in general, for concreteness, we focus on time-reversal invariant gapped systems in 1D and 3D (spatial dimensions), with $U(1)$ charge conservation symmetry. The models we consider are doubled versions of the usual topological insulator minimal Dirac models in those dimensions. 
Because of the doubled degrees of freedom, we find that the natural pumping parameter spaces in these models can display gapless singular lines due to a competition between incompatible gapping parameters (mass terms). 
The resulting pumping transport and responses are then found to depend on which singularities are encircled during the adiabatic cycles, i.e., they depend on the path and not just the end points. 
 We approach our analysis by several different methods, all with matching results.
Of note, for one of our methods we employ the Ma\~{n}es-Bardeen form of the Wess-Zumino action \cite{Witten,BardeenZumino, Manes} 
to compute the action change upon adiabatic transformations of the Hamiltonian, which is a powerful technique that has not been applied in this context.

The models on which we focus are essentially doubled versions of 1D and 3D $\mathbb{Z}_2$ topological insulators. As such they are trivial insulators according to the 10-fold classification table\cite{Schnyder_PhysRevB.78.195125, KitTable}. Indeed, the doubling allows extra possible mass terms that can be chosen to gap any  surface states without breaking the protecting discrete symmetries. Thus, while the pumping processes we consider are stable, the topological phenomena and connection to the related electromagnetic responses are not reliable since the parent topological insulator phases are rendered trivial. The response physics of the models we describe, however, may have relevance in the context of crystalline topological insulators \cite{Hsieh2012_cryst,LiangFuOctetCryst}, in which case the instabilities will be removed by requiring the preservation of, e.g., mirror symmetries. 

The paper is organized as follows. 
To illustrate the basic ideas, we start in Section \ref{sec:pump1d} with a review of adiabatic quantum pumping in 1D, considering the example of the spinless Su-Schrieffer-Heeger model\cite{SSH, ThoulessPump}. 
In Section \ref{sec:generalpump}, we proceed to explain the important concepts for Thouless pumping processes in our higher dimensional parameter spaces. 
Section \ref{sec:1+1D} is devoted to our first example in 1D. 
We consider a spinfull Su-Schrieffer-Heeger chain and demonstrate the details of the generalized adiabatic pumps through a four-fold analysis. 
First by a perturbative continuum field theory computation, second in a microscopic lattice picture and, third, in a non-perturbative approach using the Ma\~{n}es-Bardeen formula for the anomaly. 
Finally, we employ a bosonization approach to further solidify the results. 

Next, in Section \ref{sec:3+1D}, we consider a second example of a 3D time-reversal invariant insulator. In this case we again consider the
perturbative field theory computation first. We then proceed to understand the problem from the point of view of the surface state properties in the presence of a magnetic field. 
Finally, although the bosonization picture is not available, we also consider a non-perturbative  approach from the Ma\~{n}es-Bardeen form of the chiral anomaly. We then conclude with a discussion of future directions. In the appendices we add some details of the perturbative calculations.

\section{Adiabatic Quantum Pumping in 1D \label{sec:pump1d}}

We begin our work with an in-depth discussion and review of the 1D insulator example described in the introduction. 
We take a 1D insulator coupled to external, adiabatic perturbations represented by a set of $n$ parameters $\{\theta_i\}$ which enter the Bloch Hamiltonian $H(k,\{\theta_i\})$, parametrized by the crystal momentum $k$ for translationally invariant systems. 
$H(k)$ must include a minimum of two bands if it is to represent an insulator; for all of the models in this article there will be an even number of bands.

The adiabatic deformation condition implies that for all values of the parameters in the set $\{\theta_i\},$ the energy gap between these bands remains open. In this framework, if a closed path is traversed in the $\{\theta_i\}$ parameter space, then a \emph{quantized} amount of electric charge will be pumped from one side of the 1D wire to the other\cite{ThoulessPump}. While this statement applies in quite general settings, it is important to note in passing that it fails if electric charge is not conserved during the adiabatic process. 
For now we assume strict charge conservation, and as such, we would discover that during a cyclic adiabatic process, an integer number of electric charges (possibly zero) would be transferred through the sample.

For most insulators, and most choices of adiabatic perturbation cycles, the amount of transferred charge vanishes.
To find cases when the charge transferred is non-zero we need to define a few quantities:
(i) the periodic part of the Bloch functions $\vert u_{\alpha}(k,\{\theta_i\})\rangle$
which are the eigenstates of $H(k,\{\theta_i\})$ in band $\alpha$,
(ii) the adiabatic connection
${\cal{A}}_{\mu}^{\alpha\beta} = -i\langle  u_{\alpha}(k,\{\theta_i\})\vert \partial_{\mu}\vert  u_{\beta}(k,\{\theta_i\})\rangle$
where $\alpha$ labels the band indices
and $\partial_0=\partial_k,\, \partial_i = \partial_{\theta_i};$ and, finally, (iii) the first Chern number
\begin{eqnarray}
C_1&=&\frac{1}{4\pi}\int_{BZ} dk \oint_{C}d\theta^{i} {\rm{Tr}}\left[{\cal{F}}_{0i}  \right]\\
{\cal{F}}^{\alpha\beta}_{\mu\nu}&=& \partial_{\mu} {\cal{A}}^{\alpha\beta}_{\nu}-\partial_{\nu} {\cal{A}}^{\alpha\beta}_{\mu}+i [{\cal{A}}_{\mu},{\cal{A}}_{\nu}]^{\alpha\beta}
\end{eqnarray}
where the $k$ integral is over the Brillouin zone (BZ). 
The $\theta^{i}$ integral is over the 1-dimensional curve $C$ traversed during the adiabatic process in the $n$-dimensional $\{\theta_i\}$ space,
and ${\cal{F}}^{\alpha\beta}_{\mu\nu}$ is the Berry curvature of the occupied bands. 
Only in the special case when the Chern number (which is an integer by definition) is non-zero is there a finite charge pumping equal to $\Delta Q = eC_1.$

To explicitly illustrate charge pumping we will use the canonical model, i.e., the spinless Su-Schrieffer-Heeger model\cite{SSH}. 
This is a non-interacting dimerized chain with two atoms per unit cell labeled by A/B. The Hamiltonian for the electrons in such a lattice can be written
\begin{equation}
H_{1}=-t\sum_{j} c^{\dagger}_{jA}c^{\phantom{\dagger}}_{jB}-t'\sum_{j}\left(c^{\dagger}_{j+1A}c^{\phantom{\dagger}}_{jB}+{\rm{h.c.}}\right)
\end{equation}
and can be supplemented by an onsite energy term for each atom
\begin{equation}
H_{\Delta}=\Delta\sum_{j}\left(c^{\dagger}_{jA}c^{\phantom{\dagger}}_{jA}-c^{\dagger}_{jB}c^{\phantom{\dagger}}_{jB}\right).
\end{equation}
The Bloch Hamiltonian of $H_1+H_\Delta$ (i.e., the Rice-Mele model\cite{RiceMele}) is
\begin{equation}
H_{1}(k)=t'\sin k\; \tau^y +(t+t'(\cos k-1))\tau^x +\Delta\tau^z,
\end{equation}
in a basis $(c_{Ak},c_{Bk})^T$. Notice that the onsite energy Hamiltonian $H_\Delta$ breaks both inversion (P=$\tau^x$) 
and charge conjugation (C=$\tau^z K$, with $K$ the complex conjugation operator) symmetries, 
and that both of these symmetries quantize the polarization and must be broken in order for one to continuously pump charge.
Now, suppose that we have some external control over the parameters of this model which we parameterize with an angular phase $\theta$ as a curve in a two-dimensional space $(t,\Delta)$: 
  $t\equiv m\cos\theta$ and $\Delta \equiv m\sin\theta.$ 
We will adjust $\theta$ such that the perturbations are always adiabatic. In principle, this angle can vary as a function of position and/or time, and the system will respond accordingly. Linear response theory (the Kubo formula) then dictates that the corresponding current density is equal to the time derivative of the \textit{charge polarization} of the 1D system,
\begin{equation}
J_{x}=\frac{\partial P\left(\theta\right)}{ \partial t}.
\end{equation}
Meanwhile, by the continuity equation, the charge density becomes
\begin{equation}
\rho=-\frac{\partial P\left(\theta\right)}{dx}.
\end{equation}
The charge polarization for our system is given by the solid angle subtended by the curve 
$\mathbf{d}\left(k\right)=\left(t^{'}\cos k+m\cos\theta,\, t^{'}\sin k,\, m\sin\theta\right)$ which, for $m \ll t^{'} $, gives simply $P\left(\theta\right)\simeq e\theta/2\pi$. 
In fact, in this limit it is easy to see that as $\theta\to\theta+2\pi n$ then $n$ charges are pumped.

Let us now review the connection to the electromagnetic response of 1D inversion\cite{Hughes2011,Turner2010} symmetric insulators\cite{QHZ2008}.
We have seen from the explicit calculations above that charge density is bound to spatial variations of $\theta,$ and charge currents flow in response to a time-dependent $\theta.$ This topological response is identical to the Goldstone-Wilczek response\cite{goldstone1981} and is captured by a $\theta$-term effective action
\begin{eqnarray}
S_{eff}[\theta,A_\mu]&=&\frac{e}{4\pi}\int dt dx\; \theta(x,t)\epsilon^{\mu\nu}F_{\mu\nu}\nonumber\\
&=&\frac{e}{2\pi}\int dt dx\; \theta(x,t)E_{x}(x,t)\label{eq:gwaction}
\end{eqnarray}
where $F_{\mu\nu}$ is the electromagnetic field-strength tensor, and $E_{x}$ is the $x$-component of the electric field.
To see how the response is encoded, one takes the functional derivative
\begin{align}
&\langle j^{\mu}\rangle= \frac{\delta S_{eff}}{\delta A_\mu}=-\frac{e}{2\pi}\epsilon^{\mu\nu}\partial_\nu \theta(x,t),\\
&\langle \rho\rangle=-\frac{e}{2\pi}\partial_x \theta(x,t),
\nonumber\\ &\langle J^{x}\rangle=\frac{e}{2\pi}\partial_t\theta(x,t).\nonumber
\end{align}
We recognize these equations from 1D electromagnetism, and we can identify $e\theta(x,t)/2\pi$ as the electric charge polarization of the 1D insulator.

We would like to compare the \emph{difference} in the electromagnetic response of a trivial insulator and a topological insulator. 
A trivial insulator has an integer charge polarization and hence, for a homogeneous system, $\theta=2\pi n$ for some integer $n.$
On the other hand, a topological insulator will have a contribution from a half-integer polarization: $\theta=2\pi (q+1/2)$ for some integer $q.$
If we take the end points of our adiabatic path to be these two symmetry-preserving systems, then the topological response in this case turns out to be the difference\cite{QHZ2008}
\begin{eqnarray}
S_{top}&=&S_{eff}[2\pi (q+1/2),A_{\mu}]-S_{eff}[2\pi n,A_{\mu}]\nonumber\\
&=&S_{eff}[2\pi(q-n+1/2),A_{\mu}].
\end{eqnarray} 
The integer contribution $2\pi(q-n)$ to the bulk polarization can be removed by stacking the system up with other trivial insulators, and hence the topological response is due to the $\theta=\pi$ contribution. Indeed, for systems with inversion symmetry, $\theta$ is quantized to be a multiple of $\pi,$ and there is a bulk $Z_2$ topological invariant that distinguishes even (trivial) and odd (non-trivial) multiples of $\pi$\cite{QHZ2008,Hughes2011,Turner2010}.

We also see one more interesting feature. 
Since the response equations depend on spatial gradients of $\theta,$ there is some consequence of gapped \emph{spatial} evolution (as opposed to adiabatic time evolution), i.e., a gapped interface or boundary where $\theta$ varies with position. 
The response equations dictate the quantum numbers bound to regions of the system where $\theta$ is changing in space and the system remains gapped. 
For the present example, the gapped interpolation between the interior and exterior of a material where $\theta$ changes determines the amount of bound electric charge at the interface. 
This is equivalent to the usual boundary charge theorem for polarized insulators and the result of Su, Schrieffer and Heeger, (and Jackiw and Rebbi earlier\cite{jackiw1976}) that solitons in their (spinless) 
model bind localized half-charges.

To summarize, we have seen that closed adiabatic paths can lead to a pumping process, while open, symmetry-breaking paths with symmetric end points can serve to determine topological (electromagnetic) responses. Finally, these considerations also determine bound charges/states in regions where the adiabatic parameters vary in space.

\section{Thouless Pumps with Line Singularities \label{sec:generalpump}}
We now consider an enlarged parameter space. We will be as general as possible to illustrate how the reasoning works in systems as arbitrary as possible. 
The subsequent sections will discuss specific examples.

So far the effective pumping parameter space has been a two-dimensional plane parametrized by polar coordinates $m$ and $\theta$ with the origin $m=0$ removed. 
Every other point besides the origin represents a gapped Hamiltonian. Additionally, in this parameter space, only the $x$-axis, with $\theta=0$ or $\pi$, represents Hamiltonians obeying charge conjugation and inversion symmetries. 
We now enrich the parameter space by doubling the number of fermionic degrees of freedom. 
In 1D insulators, a simple model for such a system consists of a spinfull Su-Schrieffer-Heeger (SSH) chain, built from two copies (four internal degrees of freedom) of the previously discussed model. 
In 3D, a simple model  consists of a time-reversal invariant 3D gapped Dirac model with an additional doubling (eight internal degrees of freedom). Hence, it represents two copies of a usual 3D topological insulator.
Both doubled systems preserve the symmetries of the original--prior to doubling--ones.

Since these systems can be described by Dirac Hamiltonians, let us consider then for concreteness a general Dirac-type insulator described by the Bloch Hamiltonian
\begin{equation}
H\left({\textbf{k}},m,m_{5},\Delta\right).\label{eq:Hparam}
\end{equation}
It depends on momenta ${\textbf{k}}$ and
three parameters which we call $m$, $m_{5}$ and $\Delta$; this exhausts the
scenarios that we consider in this article. The adiabatic parameters control the gap in the system and dictate the possible classes of adiabatic paths.
%%General arguments that apply beyond the simple Dirac model define their interplay, as explained below.

To be explicit,  we take $m$ to represent the usual ``topological insulator gap'', or TI mass. It conventionally preserves any requisite symmetries, and will act diagonally in the two fermion subspaces resulting from the doubling. In the presence of
the protecting discrete symmetries, its value (or sign in a continuum model) effectively defines the topological regime of the model.
We can use the TI mass to define the reference points, namely $(m,m_5,\Delta)=\left(m,0,0\right)$
and $\left(-m,0,0\right),$ where by convention $m>0.$  
For our regularization convention, the former will be a symmetry-preserving trivial insulator. 
The latter would usually represent a symmetry preserving topological insulator if the system were not doubled; the doubling trivializes any $Z_2$ topological invariants. 
We note that a continuous evolution between the reference points \emph{in a straight line} in parameter space, flipping the sign of $m$,  has a vanishing gap at $m=0$.

The mass term $m_{5}$ is chosen to break \emph{all} of the symmetries that would protect a topologically non-trivial (short-range entangled) phase. Hence, one can always find a gapped interpolation between the two reference points even when constrained to the 2D parameter space $(m,m_5)$. To be concrete, we choose $m_{5}$  to act identically within each copy of the doubled system as the usual axial mass. 
The symmetry it breaks is dimension dependent, e.g., in 1D it breaks charge-conjugation and inversion, and in 3D it breaks time-reversal and inversion. 
This is the usual symmetry-breaking mass term we considered before, when defining the charge-pumping operation in the 1D spinless SSH model. 

Additionally, since we have two copies of a $Z_2$ TI--thus a topologically trivial system--the remaining mass $\Delta$ can be chosen such that it may or may not preserve the protecting symmetries; either way one can still adiabatically connect the two reference points, now in the generic 3D parameter space $(m, m_5, \Delta)$. 
We will choose $\Delta$ such that it couples the two copies. Doubling a Dirac model always introduces three distinct possibilities for such a mass term. For our analysis it will be always chosen as to break the protecting symmetry, and possibly other discrete anti-unitary symmetries not required for topological stability. For example, in our 1D model it is chosen to break both charge-conjugation, and it breaks time-reversal symmetry, the latter not required for topological stability.

As constructed, both of these masses are \emph{compatible} with $m$, 
i.e., their presence enhances the insulating gap (the corresponding mass matrices anti-commute with the TI mass). They, however,
\emph{compete} among themselves. Physically this means that if both are non-vanishing, there exist certain regions
of the parameter space such that their contributions to the gap cancel each other. Mathematically, it means that the mass matrices \emph{commute} with each other.

For a non-interacting, translation invariant Dirac system, this can be seen explicitly by looking at the spectrum at ${\textbf{k}}=0$, with both masses included:
\begin{equation}
E\left(\mathbf{k}=0\right)=\pm\sqrt{m^{2}+\left(m_{5}\pm\Delta\right)^{2}}.
\end{equation}
Clearly, when $m_5=\Delta$, the gap reduces to the TI mass contribution alone, which always vanishes at some point along the interpolation path. 
In practice, the competing mass terms introduce new singular lines into the 3D parameter space in the $m=0$ plane when $m_5=\pm \Delta.$

These lines will play an important role below, but let us briefly mention what the result would be if they were not there. If both masses were compatible with each other, instead of competing, then the spectrum would be
\begin{equation*}
E\left(\mathbf{k}=0\right)=\pm\sqrt{m^{2}+m_{5}^2+\Delta^{2}}.
\end{equation*} 
This is only singular at the origin $(0,0,0).$ 
Since any closed path in this 3D space is contractible, 
there cannot be any non-trivial, quantized pumping processes (assuming that there is no enforced symmetry that forces the path to lie fixed in a plane). 
Since each path with end points at our two reference points can be continuously deformed to all others with the same end points, 
we would also expect that every path would determine the same difference in the response between the two reference states. 
Thus, if all three mass terms were compatible, 
the properties determined by closed paths would be trivial and the response would not depend on the choice of an open path.

\begin{center}
\begin{figure}[t!]
\begin{centering}
\includegraphics[scale=0.35]{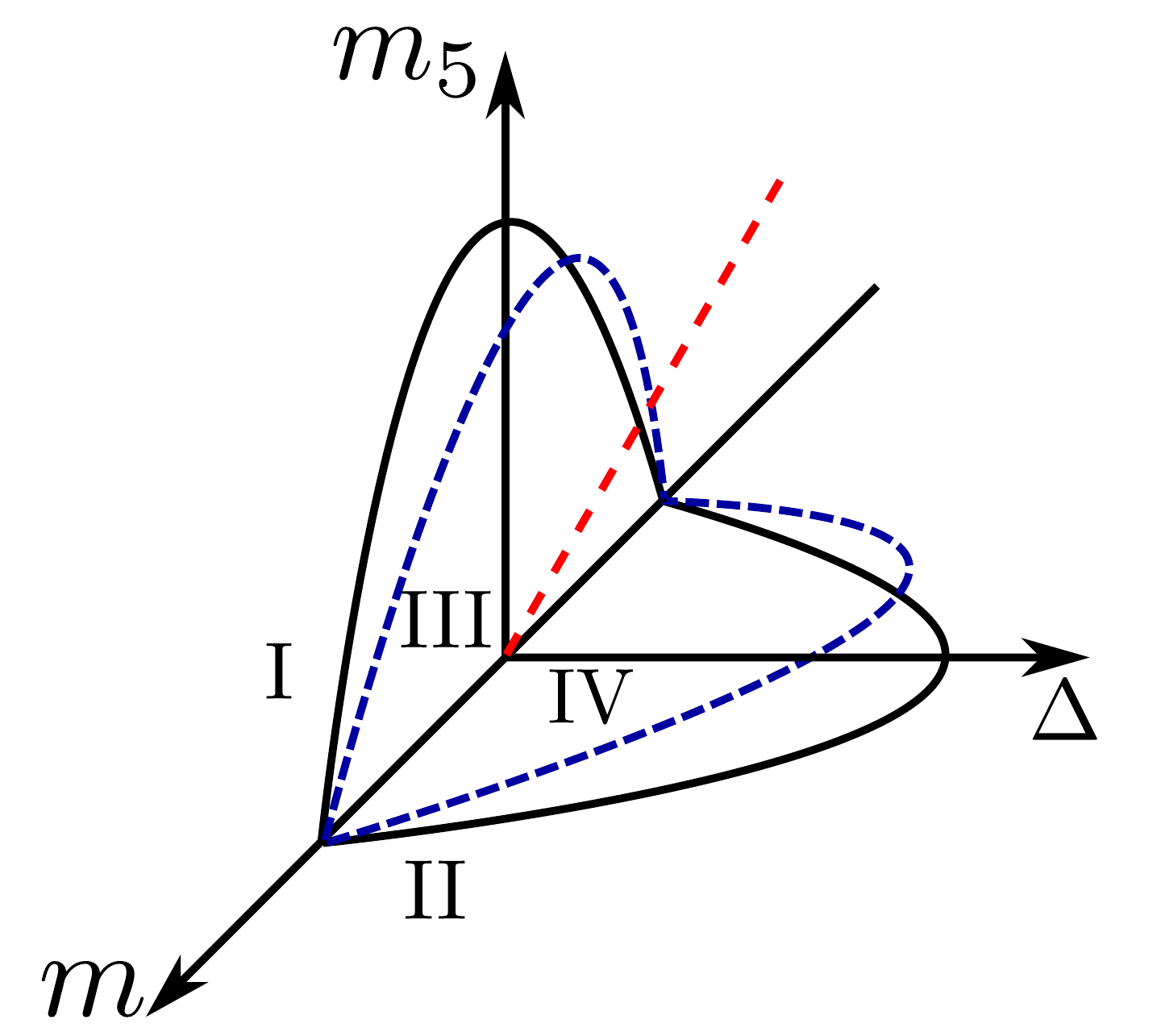}
\par\end{centering}
\caption{\label{fig:Hamilt_space}
General Hamiltonian parameter space and open path interpolations. One goal of this work is to
 compare the electromagnetic responses between points $\left(m,0,0\right)$
and $\left(-m,0,0\right)$. Interpolation paths are chosen such that
the Hamiltonian is always gapped. Trajectories $\mathrm{I}$ and $\mathrm{II}$
lie in the $mm_{5}$ and $m\Delta$ planes, respectively. The trajectories
$\mathrm{III}$ and $IV$ correspond to ``tilting'', or small deviations
of $\mathrm{I}$ and $\mathrm{II}$ away from their former planes. The red
dashed line is the ``singularity line'' $\Delta=m_5$. 
}
\end{figure}
\par\end{center}

\begin{figure}[t!]
\includegraphics[scale=0.5]{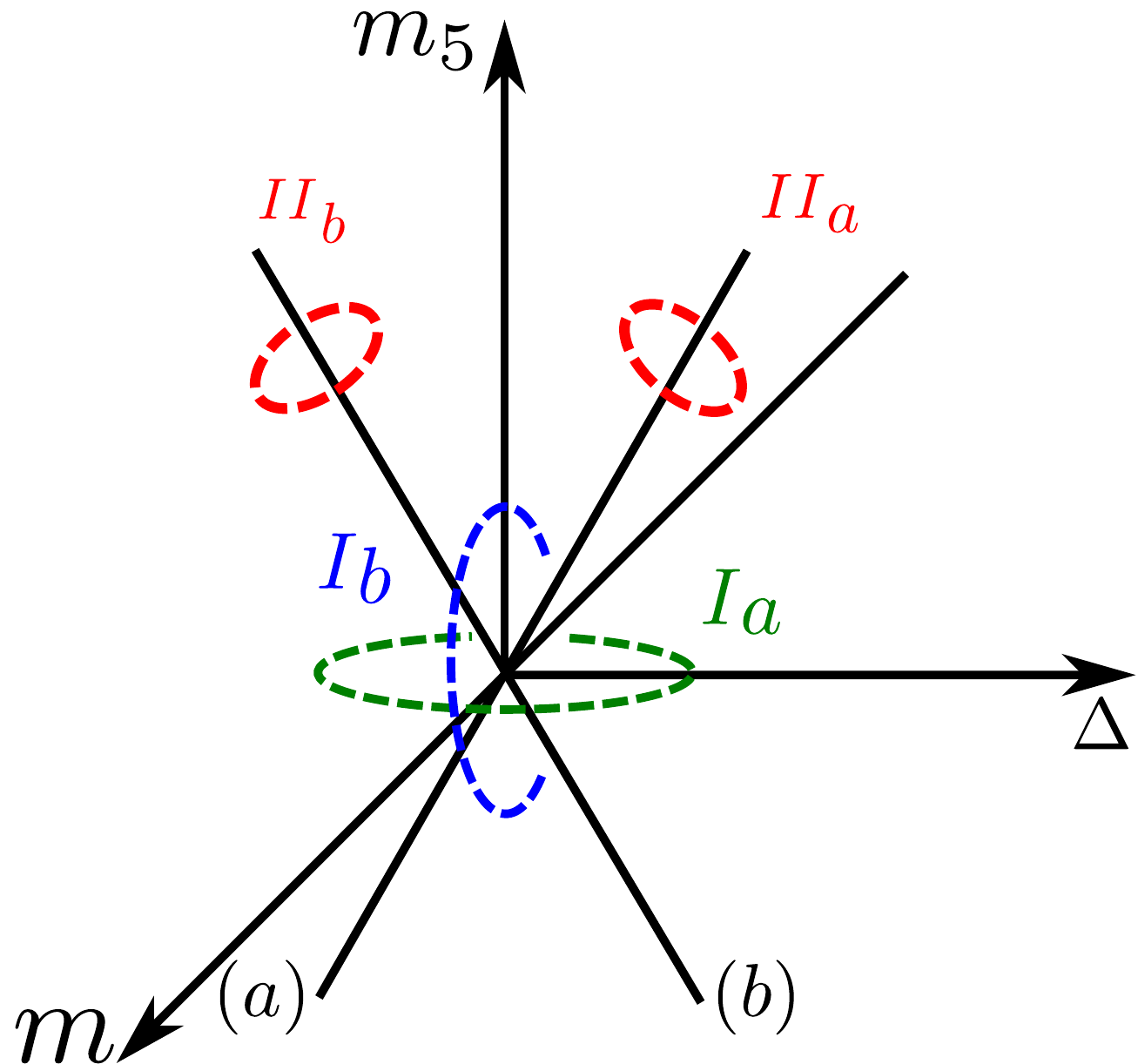}
\caption{
Adiabatic evolution closed loops in parameter space. Masses $m_{5}$ and $\Delta$
compete generating the gapless lines $\left(a\right)$ $m_{5}=\Delta$
and $\left(b\right)$ $m_{5}=-\Delta$ in the $m=0$ plane.  Evolving the Hamiltonian
around each of the represented loops gives a different pumping effect
on the Hamiltonian. 
The green and blue dashed ellipses encircle pairs of gapless lines. Trajectories can be decomposed
in terms of the paths of Fig.\ \ref{fig:Hamilt_space}.
Loops $IIa$ and $IIb$ encircle an odd number of singular lines and can be decomposed
in terms of half a horizontal path and half a vertical one. Loops $Ia$ and $Ib$, however, involve going fully around
horizontal or vertical cycles encircling an even number of gapless lines.
\label{fig:Berry-phase}}

\end{figure}

Let us return to the case of interest with competing mass terms. 
As mentioned in the previous section, 
one is able to determine the difference in the topological electromagnetic response between the two phases of the single 1D SSH chain 
by comparing the effective actions at the two reference points. 
The situation is more interesting with the higher-dimensional parameter space. 
In general, to adiabatically extract the electromagnetic response difference between the reference states, 
we must take paths between them which penetrate into the discrete-symmetry-breaking regions of parameter space. 
In order to unambiguously define charge transfer these paths must also preserve the corresponding continuous symmetries responsible for conserving the relevant charges along the pumping path.

In Fig.\ \ref{fig:Hamilt_space} we illustrate some relevant scenarios in our higher dimensional space. 
They consist of the horizontal and vertical trajectories I and II, in the $(m, m_5)$ and $(m, \Delta)$  planes, respectively. 
Trajectories III and IV correspond to tilting the previous trajectories I and II towards the singular line $m_5=\Delta$. 
The standard procedure to obtain the response in 1D and 3D is a generalized Goldstone-Wilczek calculation\cite{goldstone1981}. 
One introduces an interpolation parameter $\theta$, such that 
$\left(m,m_{5},\Delta\right)\left(\theta\right)\equiv\left(m\left(\theta\right),m_{5}\left(\theta\right),\Delta\left(\theta\right)\right)$. Correspondingly, this gives rise to a Hamiltonian computed at an arbitrary point of a given trajectory in the parameter space 
\begin{equation}
H\left(\theta\right)=H\left({\textbf{k}},m\left(\theta\right),m_{5}\left(\theta\right),\Delta\left(\theta\right)\right).
\end{equation}
One's goal then becomes to compute the change in the effective action for the gauge fields coupled to the conserved charges of the original Hamiltonian, induced by changes in the Hamiltonian parametrized by $\theta$. From such effective actions, one can distil the distinct pumping phenomena, as described in Sec.\ref{sec:pump1d}.

Regarding such effective actions, it is important to consider parametrized Hamiltonians which does not pertain to a single point in the parameter space uniformly through space and time. In other words, it may be that the parametrization value may be dependent on the space-time position, $\theta=\theta(x,t)$. If one defines the reference states as $\left(m,m_{5},\Delta\right)\left(0\right)=\left(m,0,0\right)$ and $\left(m,m_{5},\Delta\right)\left(\pi\right)=\left(-m,0,0\right)$, the Hamiltonian in distinct positions of space-time may pertain to distinct reference states. Such possibility is crucial for us, in order to be able to describe domain walls between the reference Hamiltonians with opposite TI mass. Such domain walls, as discussed, can bind localized modes carrying specific quantum numbers, which one may extract also from the effective gauge theory.

The possibility of space-time fluctuations in $\theta$ introduces difficulties in our derivations. Domain walls are represented by strongly space-time varying functions $\theta$. In order to control calculations of the pumping and charge-response effects in our effective theories, even in the presence of such defects\cite{TeoKaneDefects}, we consider building up the changes in our Hamiltonian adiabatically in infinitesimal steps, all the way up to the profile $\theta(x,t)$ desired. To realize this, we introduce a path parameter $\alpha$. Considering a small variation $\alpha \rightarrow \alpha + \delta \alpha$, with $\alpha(x,t)$ varying slowly in spacetime, we can find the change in the effective action of the conserved gauge fields to first order in $\delta \alpha.$ We then combine each infinitesimal build up to form a complete functional dependence, $\theta(x,t)$. A more detailed account of this procedure is presented in the Appendix.

We have shown examples of different classes of adiabatic paths and the consequences of traversing the various classes of paths will be analyzed in detail for the chosen examples. 
Let us, however, make a few more general comments first. 
As discussed, the two reference points respect the imposed discrete symmetries and, 
naively, one would expect that the difference in the response behavior between them should only depend on the initial and final Hamiltonians. 
Yet, we will see that there is a layer of subtlety as to how the response is calculated when there are competing mass terms. 
Although, due to doubling, both reference points correspond to trivial phases, 
with respect to the 10-fold classification table, 
the information obtained by comparing states of opposite TI mass signs aids in the determination of the quantum numbers pumped in closed paths, 
as well as the degrees of freedom trapped in solitonic defects of the parametrization variable. 
These results are then relevant for both trivial and topological phases.

Finally, the interpolation between reference Hamiltonians can be extended to adiabatic cycles (relevant both in trivial and topological systems since pumping processes do not depend on the starting Hamiltonian). 
Some generically different classes of adiabatic cycles arise, as described in Fig.\ \ref{fig:Berry-phase}. 
In the doubled 1D case, for example, the open paths $I$ and $II$ of Fig.\ \ref{fig:Hamilt_space} constitute transfers of  spin-1/2 
and a full unit (due to doubling) of electronic charge respectively. 
To constitute a true pumping (with matching initial and final ground state subspaces), 
one may stitch these paths together, resulting in the loop $II_a$ of Fig.\ \ref{fig:Berry-phase}, in which the quantum numbers of a ``full electron'' are pumped. 
The closed trajectories $I_a$ and $I_b$ then correspond to pumping of \emph{pairs} of unit-charges and spin-1/2 moments, respectively. As discussed, trajectories $II_{a,b}$ combine the separate fractional pumping, with respect to the microscopic contents of a unit cell, of a single charge and a unit of spin-1/2, thus constituting the ``full electron'' pump when traversed.

In what follows we will discuss in detail the above results, using specific examples in (1+1)D and in (3+1)D.

\section{1+1D Doubled SSH Model\label{sec:1+1D} }

\subsection{Generalities\label{sub:1DGeneralities}}

To illustrate a generalized Thouless pump in (1+1)D we will use two
copies of the SSH model by adding a second copy with equal Hamiltonian.
For definiteness, we will consider the two copies to represent fermions with up and down flavors of spin. The Hamiltonian
in the continuum limit can be written as follows
\begin{eqnarray}
\mathcal{H}_{1D} & = & \int dx\Psi^{\dagger}H_{1D}\Psi\nonumber \\
H_{1D} & = & \Gamma_{1}\left(p_{x}-eA_x\right)+\Gamma_{2}m+\Lambda_{5}m_{5}+\boldsymbol{\Delta}\cdot\boldsymbol{\Lambda},\label{eq:doubSSH}
\end{eqnarray}
in the basis $\Psi^{T}=\left(\begin{array}{cccc}
\psi_{A\uparrow} & \psi_{B\uparrow} & \psi_{A\downarrow} & \psi_{B\downarrow}\end{array}\right)$, 
with $\psi_{A\sigma},\,\psi_{B\sigma}$ fermionic annihilation operators of spin $\sigma$.
In Eq. (\ref{eq:doubSSH}), $A_x$ is the electromagnetic vector potential,
$m$ is the mass generated by a staggered hopping between the lattice
sites ("TI mass"), and $m_{5}$ is the mass generated by a staggered onsite energy.
The artificial doubling allows the introduction of three extra mass
terms $\boldsymbol{\Delta}=\left(\begin{array}{ccc}
\Delta_{1}, & \Delta_{2}, & \Delta_{3}\end{array}\right)$ 
(i.e., there exist three new $4\times4$ matrices which anti-commute with the kinetic part of the Hamiltonian) which couple the different spin subspaces and split their degeneracy.

We choose the Dirac matrices:
\begin{eqnarray}
\Gamma_{1} & = & \sigma_{0}\tau_{y}\nonumber \\
\Gamma_{0} & = & \sigma_{0}\tau_{x}\nonumber \\
\Lambda_{5} & = & \sigma_{0}\tau_{z}\\
\boldsymbol{\Lambda} & = & \boldsymbol{\sigma}\tau_{z},\nonumber
\end{eqnarray}
where $\sigma_{i}$ and $\tau_{i}$ are Pauli matrices in the spin
and sublattice (or orbital) Hilbert spaces, and Kronecker products are implicit. 
The matrices $\sigma_0$ and $\tau_0$ are $2\times2$  identity matrices.
In the absence of $\Lambda_{5}$ and $\boldsymbol{\Delta}$ masses, the model displays time-reversal and particle-hole symmetries, with the operators $T=i\sigma_{y}\tau_{0}K$ and $C=\tau_{z}K$, where $K$ is the complex conjugation operator. The remaining masses break these discrete symmetries in a pattern described in Table \ref{tab:1DSymmetries}.
\begin{table}[t]
\begin{centering}
\begin{tabular}{c|c|c}
\hline
Matrix & C & T\tabularnewline
\hline
{\large{$\Lambda_{5}$}} & {\large{$\times$}} & {\large{$\checkmark$}}\tabularnewline
{\large{$\Lambda_{x}$}} & {\large{$\times$}} & {\large{$\times$}}\tabularnewline
{\large{$\Lambda_{y}$}} & {\large{$\checkmark$}} & {\large{$\times$}}\tabularnewline
{\large{$\Lambda_{z}$}} & {\large{$\times$}} & {\large{$\times$}}\tabularnewline
\hline
\end{tabular}
\par\end{centering}
\caption{Symmetry properties of the masses for the (1+1)D doubled SSH model under particle-hole $\left(C\right)$
and time-reversal $\left(T\right)$ transformations. The corresponding symmetry is broken/preserved for $\times$/$\checkmark$ marks, respectively \label{tab:1DSymmetries}}
\end{table}
The commutation relations,
\begin{eqnarray}
\left\{ \Gamma_{a},\Lambda_{j}\right\} =\left\{ \Gamma_{a},\Lambda_{5}\right\}  & = & 0\nonumber \\
\left\{ \Lambda_{i},\Lambda_{j}\right\}  & = & 0\label{eq:commuts}\\
\left[\Lambda_{5},\Lambda_{j}\right] & = & 0,\nonumber
\end{eqnarray}
with $a=0,\,1;\ i,\, j=x,\, y,\, z$, imply that, although all masses are compatible with the TI mass $m$, the mass
$m_{5}$ and the set $\boldsymbol{\Delta}$ actually compete. In the absence of $m$,
this can lead to gapless spectra when  $|m_5|=|\boldsymbol{\Delta}|.$ 

Let us consider the effective electromagnetic action and quantum pumping processes in this system. First, let us look at the situation with $\boldsymbol{\Delta}=0,\, m_{5}=0$. This
corresponds to just a pair of disconnected linear chains, and each chain is known to have topologically
distinct phases for $m>0$ and $m<0$. To find the topological response from each independent chain,
one interpolates between these phases using $H\left(\alpha\right)$,
with $\alpha$ an angle varying slowly in space-time as discussed in Sec.\ \ref{sec:generalpump} (the complete, doubled chain, system is trivial overall and will give a trivial total response).
In order to achieve an adiabatic interpolation, the gap must not close as $\alpha$ is varied. Hence, we must avoid the point $m=m_5=\Delta=0$ and the lines $m=0, m_5=\pm \Delta.$ From this parameterization we can see that this model contains all the features discussed in Sec.
\ref{sec:generalpump}.

As an example of a class of adiabatic paths, consider the following interpolation
\begin{equation}
H_{\mathrm{I}}\left(\alpha\right)=\Gamma_{1}\left(p_{x}-e A_x \right)+\Gamma_{0}m\cos\alpha+\Lambda_{5}m\sin\alpha,\label{eq:1DChiralPath}
\end{equation}
with $\alpha$ changing from $0$ to an arbitrary angle $\theta$. 
This corresponds to the ``pure'' axial-mass path, labeled path $\mathrm{I}$ in Fig.\ \ref{fig:Hamilt_space}.
The fermionic path integration for the change in effective action due to an infinitesimal $\alpha$, described in detail in Appendix \ref{app:Appendix-A}, 
is nothing but a Goldstone-Wilczek-like calculation and gives rise to an axion term. 
Building up the contributions of infinitesimal changes in $\alpha$ from $0$ to an arbitrary finite angle $\theta$, one finds
\begin{equation}
S_{eff}^{\mathrm{I}}=2\Theta_{1D}\label{eq:1DpathI}
\end{equation}
where $\Theta_{1D}$ is the standard 1D  "$\theta$-term"
\begin{equation}
\Theta_{1D}=\frac{e}{2\pi}\int dxdt\theta E.
\end{equation}
This action dictates both the adiabatic pumping process (by considering full evolutions of $\theta$ by $2\pi$), and quantum numbers bound to solitons, like domain-walls in the TI mass. In particular, the latter case can be described by $\theta(x,t)=\pi \Theta(x)$ where $\Theta(x)$  is the Heaviside step function. In practice, this implies that each wire contributes 
charge $\pm e/2$ at the ends of the system, yielding in total an integer charge bound to the edge (and likely unobservable in realistic systems because of the charge polarization ambiguity in lattice models\cite{vanderbilt1993}).  %Thus, the system harbors a topological response albeit with a trivial (probably unobservable) value.

One would naively expect the response difference between reference Hamiltonians at $\theta(x,t)=0$ and $\theta(x,t)=\pi$  to be a property associated only with the initial and final Hamiltonians $H\left(0\right)$
and $H\left(\pi\right)$. The path taken in the parameter space, however, can be easily seen to be relevant in our model. For example, considering the
path $\mathrm{II}$ in Fig.\ \ref{fig:Hamilt_space} (for $\boldsymbol{\Delta}=\Delta_{z}\hat{z},$
say), one lets $\Delta_{z}=m$ and defines,
\begin{equation}
H_{\mathrm{II}}\left(\alpha\right)=\Gamma_{1}\left(p_{x}-e A_x \right)+\Gamma_{2}m\cos\alpha+\Lambda_{z}m\sin\alpha.
\end{equation}
Following the same Goldstone-Wilczek type of calculation as before, one finds for the effective action
\begin{equation}
S_{eff}^{\mathrm{II}}=0\label{eq:1DpathII}
\end{equation} which is clearly different from Eq. \ref{eq:1DpathI}.

If the initial and final Hamiltonians $H\left(0\right)$ and
$H\left(\pi\right)$ are the same, we would expect that any continuous deformation of the path in
parameter space should result in the same response difference for the two cases. The discrepancy
between (\ref{eq:1DpathI}) and (\ref{eq:1DpathII}) can only arise from the fact
that the masses $m_{5}$ and $\Delta_{z}$ compete.
This is so as one cannot deform path $\mathrm{I}$ to $\mathrm{II}$ without passing through a gapless point/line. 
Further support for this idea can be found by calculating the response difference on the tilted paths $\mathrm{III}$ and $\mathrm{IV}$ 
in Fig.\ \ref{fig:Hamilt_space}. One can do this by considering, for example,
\begin{eqnarray}
H_{\mathrm{III}}\left(\phi\right) & = & \Gamma_{1}\left(p_{x}-e A_x \right)\\
 &  & +\Gamma_{0}m\cos\alpha+\Lambda_{5}m\sin\alpha+\Lambda_{z}\Delta_{z}\sin\alpha\nonumber
\end{eqnarray}
and computing corrections in the effective action perturbatively in $\Delta_{z}/m$.   
As long as this dimensionless parameter is small,
we do not cross the gapless line. The procedure is described in detail in Appendix \ref{app:Appendix-A}. An analogous calculation can be done for path $\mathrm{IV}$, perturbatively in $m_{5}/m$. What one finds is that the effective
actions do not change from their respective original behaviors, i.e.,
\begin{eqnarray}
S_{eff}^{\mathrm{III}}=S_{eff}^{\mathrm{I}} & = & 2\Theta_{1D}\nonumber \\
S_{eff}^{\mathrm{IV}}=S_{eff}^{\mathrm{II}} & = & 0.
\end{eqnarray}
Table \ref{fig:-Summary1D} summarizes the above results. They suggest that, indeed, the change in the behavior between paths $\mathrm{I}$ and $\mathrm{II}$ is due to the discontinuous jumps at singular lines in parameter space, as opposed to a smooth continuous evolution.

There is, however, a subtlety in this discussion of the electromagnetic response. As far as the topological response properties of lattice \emph{fermionic} systems go, $S_{eff}^{\mathrm{I}}$ is essentially equivalent to $S_{eff}^{\mathrm{II}}$ since they both represent the response properties of trivial insulators. To put this another way, the integer charge polarization from Eq. \ref{eq:1DpathI} can be removed by a gauge transformation on the Bloch wavefunctions. Closed adiabatic evolutions of the Hamiltonian, however, still lead to pumping of quasi-particles. Closing a circular path along trajectory $\mathrm{I}$ pumps $\emph{twice}$ an electric charge in the present scenario.

Remarkably, these considerations show that a closed path evolution of the Hamiltonian
around the gapless line, i.e., going forward on path $\mathrm{I}$ and reverse on path $\mathrm{II}$ will not bring the system back to its original
state, which indicates a ground state degeneracy, and a type of Berry
phase holonomy that is being gathered in this process. As a further consequence, we expect some other `hidden' charge pumping during this process. In fact, the spinfull SSH Hamiltonian allows for a set of SU(2) conserved charges,
\begin{equation}
t_{a}=\frac{1}{2}\left\{ \sigma_{x}\tau_{0},\,\sigma_{y}\tau_{0},\,\sigma_{z}\tau_{0}\right\} =S_{a}, \label{eq:spinmat},
\end{equation}
i.e., just the spin components themselves. The different $\boldsymbol{\Delta}$ masses break this SU(2) spin invariance, but an arbitrary path in parameter space through $\boldsymbol{\Delta}$ still preserves a spin-U(1) symmetry for rotations along a given axis. Fixing a single conserved component at, say, $S_z$ and gauging this symmetry by introducing a gauge field $A_{\nu}^{S_{z}}$, path II now induces a finite response. One finds, following the regular Goldstone-Wilczek calculation,
\begin{equation}
S_{eff}=\frac{1}{\pi}\int d^{2}x\epsilon^{\mu\nu}\left[\theta_{I}\partial_{\mu}A_{\nu}+\frac{1}{2}\theta_{II}\partial_{\mu}A_{\nu}^{S_{z}}\right] \label{eq:SSH_act}
\end{equation}where $\theta_{I,II}$ corresponds to adiabatic evolution of the families of Hamiltonians along the I or II trajectories. Also,  although we chose spin conservation along the z-direction, an arbitrary direction could have been chosen, with an appropriately parametrized corresponding rotation.

These results suggest that the competition of mass terms is related to the known physics of spin-charge separation in spinfull SSH chains. To further explore these phenomena, we now proceed to consider a microscopic picture of this problem, which is convenient to study the physics of edge bound modes.

\begin{figure}
\begin{centering}
\begin{tabular}{|c|c|}
\hline
Path & Response\tabularnewline
\hline
$\mathrm{I}$ & $2\Theta_{1D}$\tabularnewline
\hline
$\mathrm{II}$ & $-$\tabularnewline
\hline
$\mathrm{III}$ & $2\Theta_{1D}$\tabularnewline
\hline
$\mathrm{IV}$ & $-$\tabularnewline
\hline
\end{tabular}
\par\end{centering}

\caption{ Summary of the electromagnetic $\theta$-term responses calculated in each path of Fig. \ref{fig:Hamilt_space}
for the spin-doubled 1D system.\label{fig:-Summary1D}}
\end{figure}

\subsection{Microscopic picture \label{sub:Microscopic1D}}

We can shed some light on the above findings by considering an exactly
solvable lattice model of the continuum model above, still parameterized by $m,\, m_{5}$ and $\Delta$. The lattice
(doubled) SSH model is
\begin{eqnarray}\label{eq:SSHmic}
H_{mic} & = & \sum_{\sigma=\pm}\sum_{j=1}^{N-1}\left(1-\eta\right)\left(a_{j+1\sigma}^{\dagger}b_{j\sigma}+b_{j\sigma}^{\dagger}a_{j+1\sigma}\right)\nonumber \\
 &  & +\sum_{\sigma=\pm}\sum_{j=1}^{N}\left(1+\eta\right)\left(a_{j\sigma}^{\dagger}b_{j\sigma}+b_{j\sigma}^{\dagger}a_{j\sigma}\right)\nonumber \\
 &  & +m_{5}\sum_{\sigma=\pm}\sum_{j=1}^{N}\left(a_{j\sigma}^{\dagger}a_{j\sigma}-b_{j\sigma}^{\dagger}b_{j\sigma}\right)\nonumber \\
 &  & +\Delta_{z}\sum_{\sigma=\pm}\sum_{j=1}^{N}\sigma\left(a_{j\sigma}^{\dagger}a_{j\sigma}-b_{j\sigma}^{\dagger}b_{j\sigma}\right),
\end{eqnarray}
where $\sigma=\pm$ for up/down spins, and $\eta$ is the deviation
from the (normalized to 1) initial hopping amplitude. We have only included the $\Delta_{z}$ spin-dependent mass term, though we will mention the other spin mixing
terms later. The TI mass parameter $m$ is fixed by $\eta.$

It is enough to study the limit $\eta=-1$, beginning with $m_{5}=\Delta_{z}=0.$ At this point in parameter space, the correlation length vanishes, and the eigenstates are formed by
\begin{equation}
d_{j\pm\sigma}^{\dagger}=\frac{a_{j+1\sigma}^{\dagger}\pm b_{j\sigma}^{\dagger}}{\sqrt{2}},
\end{equation}
which leave the exact zero-modes $a_{1\sigma}$ and $b_{N\sigma}$ at the ends of the open chain. The presence
of the spin degrees of freedom from the doubling enforces additional degeneracies
in the ground state of the open chain. In this flat band limit, we can write down the
boundary theory with $m_{5}$ and $\Delta_{z}$ as perturbations, which
 simply reads
\begin{equation}
H_{edge}=\left(\begin{array}{cccc}
m_{5}+\Delta_{z} & 0 & 0 & 0\\
0 & m_{5}-\Delta_{z} & 0 & 0\\
0 & 0 & -m_{5}-\Delta_{z} & 0\\
0 & 0 & 0 & -m_{5}+\Delta_{z}
\end{array}\right),\label{eq:Hedge1d}
\end{equation}
in the basis $\left\{ \begin{array}{cccc}
\left|1_a \uparrow\right\rangle  & \left|1_a \downarrow\right\rangle  & \left|N_b \uparrow\right\rangle  & \left|N_b \downarrow\right\rangle \end{array}\right\} $, where $1_a,N_b$ represent the sites on the left/right end.

\begin{figure}
\subfloat[\label{fig:Chain1suba}]{\begin{centering}
\includegraphics[scale=0.17]{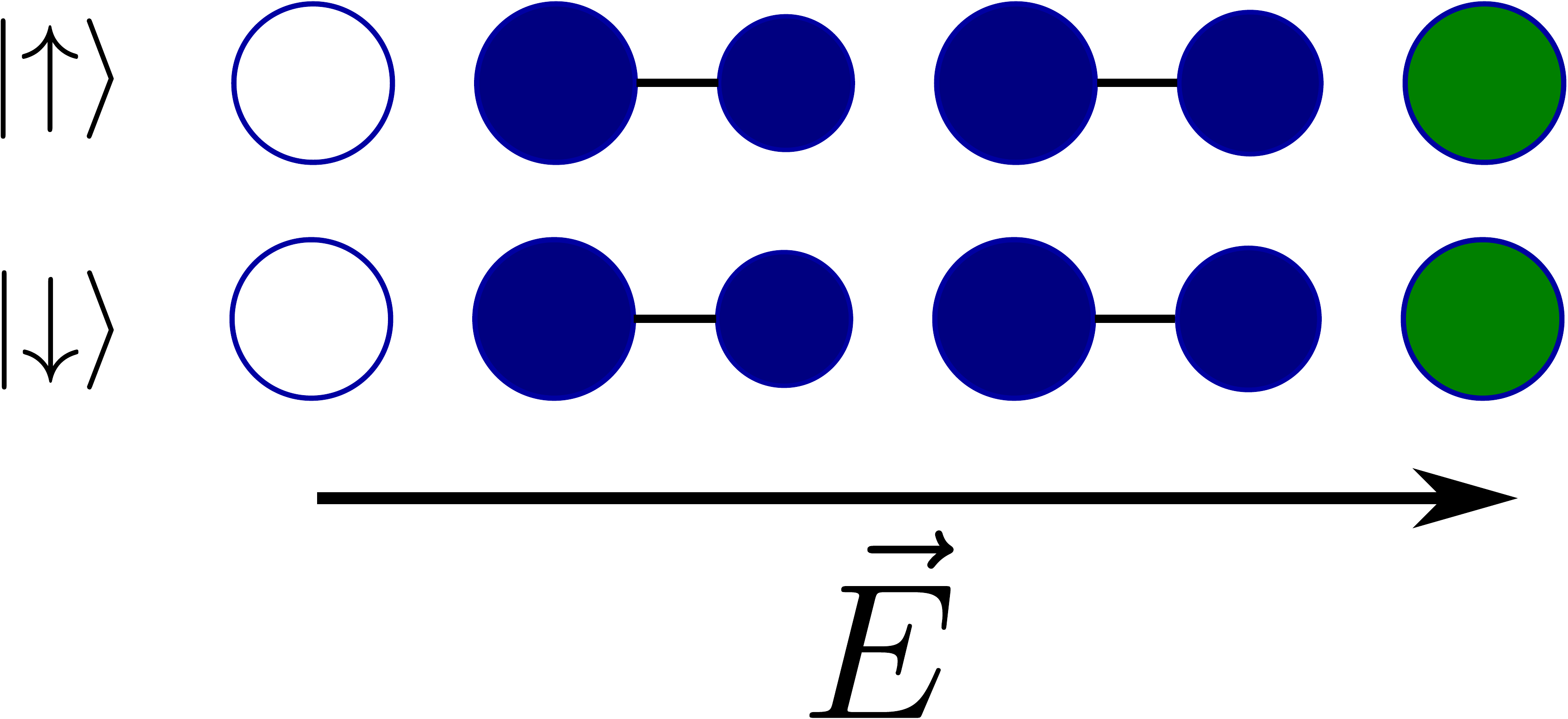}
\par\end{centering}

}

\subfloat[\label{fig:Chain1subb}]{\begin{centering}
\includegraphics[scale=0.17]{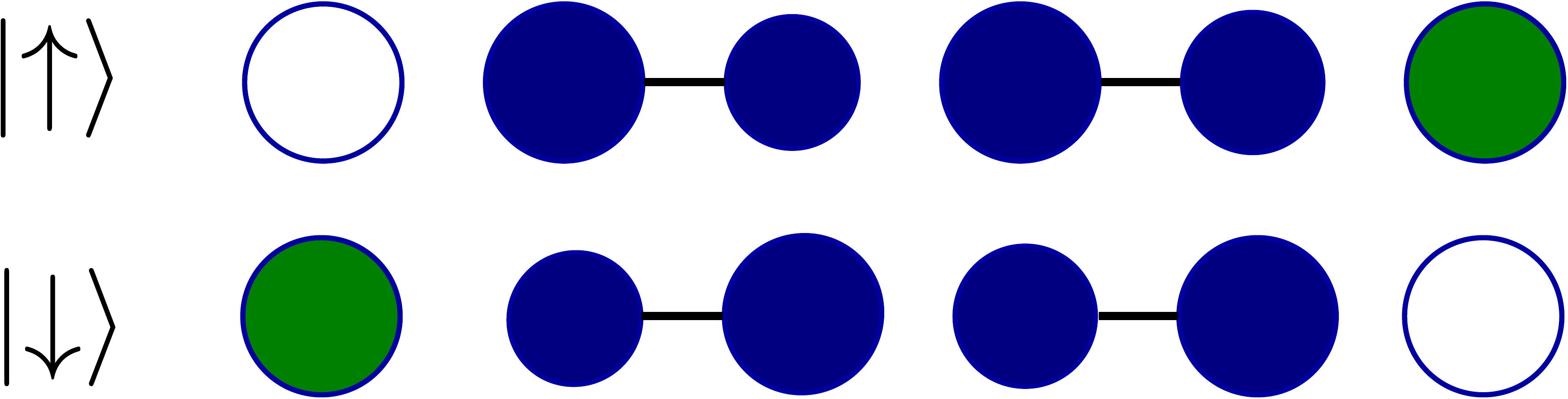}
\par\end{centering}

}

\caption{Ground states for open chains,
$\eta=-1,\, m_{5},\Delta_{z}>0$. (a) $m_{5}\gg\Delta_{z}$, one sees
that the system is polarized (b) $m_{5}\ll\Delta_{z}$, one now sees the spin polarization; Blue and
green circles correspond to filled states in the bulk and edges of
the chains respectively. Effects of finite values of $m_{5},\,\Delta_{z}$
on the site densities are represented by the relative size of the
circles. In the thermodynamic limit with $m_{5},\,\Delta\rightarrow0$,
the staggered density pattern disappear but the filled ground states
remain the same.\label{fig:Chain1}}

\end{figure}

From \eqref{eq:Hedge1d}, the scenarios of bound states implied by the axion action \eqref{eq:SSH_act} can be studied easily,
and are described pictorially in Fig.\ \ref{fig:Chain1}. At half filling, i.e., when the bulk periodic chain is insulating, two out of the four boundary modes will be occupied. When $\left|m_{5}\right|>\left|\Delta_{z}\right|$,
the lowest energy states reside on a single edge. This leads to a ground-state occupation of the states on the given edge, and
hence a non-vanishing charge polarization of the system, which then breaks C and inversion symmetries. On the other hand,
for $\left|m_{5}\right|<\left|\Delta_{z}\right|$ the ground-state occupation has one state occupied on
each edge, and for our choice of a $\Delta_z$ mass term these states have opposite spin projection in z direction. This system has no charge polarization, and hence no boundary charge, but it does have overall dangling spins at the edges.

The finite polarization state may be understood as a spontaneously symmetry broken state.
Concretely, one considers finite $m_{5}$ and $\Delta$ and fills the two lowest energy boundary modes. In the thermodynamic limit
with an infinite chain, the states on the boundaries are isolated and cannot tunnel between each end. If
we take limits $\Delta_{z}\rightarrow0$ first, followed by $m_{5}\rightarrow0$,
we are led to a spontaneously symmetry broken state with non-vanishing
charge polarization due to the gapped bulk. Inverting the order of the limits, however, leads to a state with
dangling spins at the edges.  This state presents no spontaneously broken symmetries; the only possible broken symmetry would be time reversal, however, as the two opposite spin states are degenerate at each edge, any superposition is possible for the ground state, and time-reversal symmetry is not broken. Under a spin U(1) symmetry conservation (in the z-direction, as we considered before), all that can be said is that the domain wall in the TI mass carries a single unit of spin 1/2.

Regarding the other spin mixing terms, if we consider non-zero values of $\Delta_{x}$,
$\Delta_{y},$ as well as $\Delta_{z}$, we may find a different direction
for the end state spin polarization. Nevertheless, if $m_5\to 0$ and then the $\Delta_{i}\rightarrow0$
limit is taken, each end electron forms a spin-1/2 degeneracy with
full SU(2) invariance. This also matches the result of the effective action with gauged SU(2) symmetry.

This picture is the hallmark of spin-charge separation for SSH models with
spin.\cite{Su1980, HKSS_RMP,KivelsonEFrac} Domain wall excitations in this system can be chargons or holons,
with charge $\left(\pm e\right)$, or spinons, with spin-1/2. These
results are not limited to the flat band limit, and one may verify
them analytically by considering Eq. (\ref{eq:SSHmic}) with $\eta\neq 1.$ One can simulate the edge by a domain wall with a kink
in the TI mass. The other edge is simulated by an anti-kink. Introduction
and projection of $m_{5}$ and $\Delta_{z}$ into these states recovers
the discussed results. Even though the charged kinks carry integer charge they are still fractionalized objects compared to the local degrees of freedom in the unit cell since the fundamental excitation is an electron with both charge and spin. For example, adding a `full' electron to a positively charged kink takes away its charge but converts it to a spin-1/2 kink. Thus, these fundamental kink defects will contain spin or charge, but not both. This type of spin-charge separation is also seen in flux defects in the quantum spin Hall state\cite{QiZhangSpinCharge,RanVishLee}.

The microscopic picture of spin-charge separation is linked to the two planar (non-tilted) adiabatic paths by our action Eq. \eqref{eq:SSH_act} (in the limit of spin-z conservation, for simplicity).  Suppose now we take a closed adiabatic path which encloses a singular line. We start in a symmetry preserving state with $ m_5=\Delta_z=0$ and then perform a $\pi$-rotation in trajectory I ($m-m_5$-plane) and then a $-\pi$-rotation in trajectory II ($m-\Delta_z$ plane) that brings us back to our starting point. We can ask if this system is different than, say, taking a full $2\pi$-rotation along trajectory I ($m-m_5$-plane.) If we consider these adiabatic paths as spatial evolutions/interfaces, then from the analysis of the bound quantum numbers, in the first case we would have both a bound chargon and a spinon, i.e., an additional full electron, and in the latter case we would have either two chargons or two holons. Hence the different classes of adiabatic paths give rise to physically distinct pumping
processes, which are captured by Eq. \eqref{eq:SSH_act} when considering domain-walls in the corresponding $\theta_{I,II}$ angles.

\subsection{Adiabatic Pumping and the Ma\~{n}es-Bardeen formula \label{sub:1dmanes}}

We can also consider the adiabatic evolutions of the gapped 1D Hamiltonians in a more systematic way.
Effective actions carrying topological content can be deduced with minimal labor by deploying an analysis pioneered by Witten \cite{Witten} and further developed in Refs. \onlinecite{BardeenZumino,Manes}. Given a system of fully gapped fermions, their currents are defined by variations with respect to external gauge fields via the effective action for these fields. Such effective actions may carry topological terms, which are defined by corresponding anomalies; the results of  Refs. \onlinecite{Witten, BardeenZumino, Manes} indicate how to construct these topological terms systematically, for any type of fermionic mass fields. 

To apply this method, let us re-analyze the doubled SSH model Hamiltonian \eqref{eq:doubSSH}. 
Noticing that $i\Gamma_0\Lambda_5=\tilde{\Gamma}$ defines a chiral symmetry, one may rewrite the fermion operators as $\Psi\rightarrow \chi_{R,L}=[(1\pm \tilde{\Gamma})/2] \Psi$. We will start from the most symmetric system and introduce the analysis of the pumping paths by a symmetry breaking procedure. Thus, the starting point is the action of the doubled massless relativistic fermions in 1+1D,
\begin{equation}
\mathcal{S}=\int d^2x \left[
\bar{\chi}_{L}\gamma\cdot i\left(\partial+L\right)\chi_{L}+\bar{\chi}_{R}\gamma\cdot i\left(\partial+R\right)\chi_{R}
\right].
\end{equation}
Here, $\bar{\chi}_{i}=\chi_{i}^{\dagger}\gamma^{0}$ and the  Lagrangian Dirac matrices are
\begin{eqnarray}
\gamma^0 & = & \Gamma_0 = \sigma_{0}\tau_{x}\nonumber \\
\gamma^1 & = & \gamma^0 \Gamma_1= i \sigma_{0}\tau_{z}.
\end{eqnarray}
Also, the chiral matrix reads, explicitly, $\tilde{\Gamma}=\sigma_{0}\tau_{y}$.

In general we can introduce a \emph{pair} of non-Abelian gauge fields $R,L$, which gauge all the symmetries of this model. Let us analyze this problem in depth. The fields display a vector U(1) symmetry $\chi_{L}^{'}=e^{i\theta}\chi_{L}
 $ and $\chi_{R}^{'}=e^{i\theta}\chi_{R}$ and an axial U(1) symmetry, $\chi_{L}^{'}=e^{i\tilde{\Gamma}\theta}\chi_{L}=e^{-i\theta}\chi_{L}
 $ and $\chi_{R}^{'}=e^{i\tilde{\Gamma}\theta}\chi_{R}=e^{i\theta}\chi_{R}$. Furthermore, they display an $SU(2)_{+}\times SU(2)_{-}$ symmetry. To see this, consider the following matrices
 \begin{eqnarray}
 t_{a} & = & \frac{i}{2} \tilde{\Gamma}\gamma^0 \left\{ \Lambda_{x},\,\Lambda_{y},\,\Lambda_{z}\right\} \\
 \lambda_{a} & = & \frac{i}{2}\gamma^0 \left\{ \Lambda_{x},\,\Lambda_{y},\,\Lambda_{z}\right\} .
 \end{eqnarray}
In our particular representation,
\begin{eqnarray}
t_{1}=\frac{1}{2}\sigma_{x}\tau_{0},\ 		\lambda_{1}=\frac{1}{2}\sigma_{x}\tau_{y} \\
t_{2}=\frac{1}{2}\sigma_{y}\tau_{0},\ 		\lambda_{2}=\frac{1}{2}\sigma_{y}\tau_{y} \\
t_{3}=\frac{1}{2}\sigma_{z}\tau_{0},\ 		\lambda_{3}=\frac{1}{2}\sigma_{z}\tau_{y}.
\end{eqnarray}
Notice that $\lambda_{a}=\tilde{\Gamma}t_{a}$ and that the $t_a=S_a$ matrices are nothing but the spin operators from \eqref{eq:spinmat}. The commutation relations obeyed by these matrices are $[t_{a},\tilde{\Gamma}]=[\lambda_{a},\tilde{\Gamma}]=0$, as well as
\begin{align}
&\left[t_{a},t_{b}\right]	=	i\epsilon_{abc}t_{c},
\nonumber \\
&\left[\lambda_{a},\lambda_{b}\right]	=	i\epsilon_{abc}t_{c},
\nonumber \\
&\left[t_{a},\lambda_{b}\right]	=	i\epsilon_{abc}\lambda_{c}.
\end{align}
The $SU(2)_{+}\times SU(2)_{-}$ symmetry is manifestly generated as
\begin{equation}
g_{\pm}=e^{i\left(\frac{t_{a}\pm\lambda_{a}}{2}\right)\theta^{a}}=e^{i\frac{1\pm\tilde{\Gamma}}{2}t_{a}\theta^{a}}.
\end{equation}

Now we will begin introducing mass terms, and hence start breaking the symmetries. First, we consider adding the TI mass, which breaks the axial symmetries locally and globally. In fact, this mass term also breaks the  $SU(2)_{+}\times SU(2)_{-}$ symmetry to its diagonal $SU(2)$ subgroup, $g_{+}=g_{-}=g$ and $g_{+}g_{-}=e^{it_{a}\phi^{a}}$. We thus must constrain the non-Abelian gauge fields $R=L\equiv A$ and the action reads
\begin{eqnarray}
\mathcal{S} & = & \int d^2x
\Big[\bar{\chi}_{L}\gamma\cdot i \left(\partial+A\right)\chi_{L}+\bar{\chi}_{R}\gamma\cdot i \left(\partial+A\right)
\chi_{R}
\nonumber \\
& & +m \left(\bar{\chi}_{L}\chi_{R}+\bar{\chi}_{R}\chi_{L}\right) 
\Big].
\end{eqnarray}
The terms on the second line represent the TI mass. 
The remaining symmetry of the problem is now $SU(2)\times U(1)$, and we gauge it as
\begin{equation}
A=-it_{A}A_{S}^{A}-i\mathbf{1}A_{e},
\end{equation}
where $A_S$ and $A_e$ are the spin and electromagnetic gauge fields.

It is our desire to consider transformations of the original Hamiltonian which send $m \rightarrow -m$. For this to be accomplished, we will have to consider transformation paths which break the discrete symmetries of the Hamiltonian, while conserving charges for transport. This can be achieved, in the present context, by considering
\begin{eqnarray}
\mathcal{S} & = & \int d^2x \Big[
\bar{\chi}_{L}\gamma\cdot i \left(\partial+A\right)\chi_{L}+\bar{\chi}_{R}\gamma\cdot i \left(\partial+A\right)\nonumber
\chi_{R}\\ & & +m \left(\bar{\chi}_{L} U \chi_{R}+\bar{\chi}_{R} U^\dagger \chi_{L}\right)
\Big]
\end{eqnarray}
where generically \begin{equation}
U=e^{i\left(\theta_I+\boldsymbol{\theta}_{II}\cdot\mathbf{t}\right)}.
\end{equation}

We are now set up to consider the pumping processes. The two limiting cases of interest are: (i)  path I rotations parameterized as $\theta_I: 0\rightarrow \pi, |\theta_{II}|=0,$ and (ii) path II rotations parameterized as $\theta_I: 0, \boldsymbol{\theta}_{II}=(0 \rightarrow \pi) \hat{z}$. Notice that path II breaks the SU(2) symmetry, but not fully. There is always a projection of the charges along the path which is conserved, which we fix to be $t_3=S_z$ for concreteness. We can also consider a general interpolation between these two limiting cases, as
\begin{eqnarray}
U & = & \cos\alpha+i\sin\alpha\left(\cos\phi+\sin\phi\hat{\alpha}\cdot\mathbf{t}\right) \\
  & := & \cos\alpha+i\sin\alpha\left(\cos\phi+\sin\phi t_{3}\right), \label{eq:evoU}
\end{eqnarray}
where $\hat{\alpha}$ is a unit vector pointing in the SU(2) $\mathbf{t}$ direction corresponding to a choice of a conserved U(1) subgroup. 
Hence, the interpolation between positive and negative TI masses $m$ is fixed by $\alpha:0\rightarrow \pi$, while the interpolation between paths I and II is fixed by $\phi:0 \rightarrow \pi/2$.
The gauge fields along these paths are reduced to just
\begin{equation}
A=-it_{3}A_{S_{z}}-i\mathbf{1}A_{e}.
\end{equation}
These are deduced from the conserved U(1) charges, spin and electromagnetic, and are the fields which couple to the conserved charges whose pumping we probe.

At this point, we are ready to introduce the gauged Wess-Zumino action of Ma\~{n}es-Bardeen. This action captures the topological part of the response of gapped fermions to variations in its mass-generating fields (as well as external gauge fields). Such a topological contribution is fully determined by gauge invariance and anomaly inflows, which are related to the computation of triangle graphs as those used in our previous diagrammatic analysis (for a recent in-depth analysis, see Ref. \onlinecite{StoneLopes} and references therein). This connects  Ma\~{n}es-Bardeen's  Wess-Zumino action directly to our problem at hand;  we will soon see that the method has computational advantages over the diagrammatic approach when considering general interpolation paths  between the reference points in the presence of all conserved charges gauge fields.
In 1+1D, the gauged Wess-Zumino action of Ma\~{n}es-Bardeen reads, in differential form notation,
\begin{eqnarray}
\delta S_{WZ}&=&C\int_{M_{3}}tr\left[\omega_{3}\left(R\right)-\omega_{3}\left(L\right)+\frac{1}{3}\left(U^{-1}dU\right)^{3}\right] \\
&&+C\int_{M_{2}}tr\left[dUU^{-1}L-RU^{-1}dU-RU^{-1}LU\right], \nonumber
\end{eqnarray}
where $\omega_{3}$ is the Chern-Simons form
\begin{eqnarray}
\omega_{3}\left(A\right)=tr\left[AdA+\frac{2}{3}A^{3}\right]
\end{eqnarray}
where we have restored the gauge fields $L$ and $R$ for a moment, $M_2$ is the 1+1D manifold, $M_3$ is the extended manifold for the Wess-Zumino-Witten term, and the normalization constant $C$ can be fixed, for example, by considering the usual (anomalous) chiral rotation, whose result is known, and then comparing the pre-factors. 

When considering our gapping terms we must fix $L=R=A$, and this action simplifies to
\begin{eqnarray}
\delta S_{WZ}&=&C\int_{M_{3}}tr\left[\frac{1}{3}\left(U^{-1}dU\right)^{3}\right] \\
&&+C\int_{M_{2}}tr\left[dUU^{-1}A-AU^{-1}dU-AU^{-1}AU\right]. \nonumber
\end{eqnarray}
Since the only space-time dependent function in $U$ is $\alpha$, see Eq. \eqref{eq:evoU}, one can write $dU=U_{\pi/2}d\alpha$, where $U_{\pi/2}\equiv U\left( \phi ,\pi/2+\alpha \right)$. Also, the result simplifies since the remaining gauge field components commute among themselves, and with $U$. Finally, the usual chiral anomaly in 1+1D for a doubled Dirac system implies $Ctr\left[\mathbf{1}\right]=\frac{1}{2\pi}$, where $tr[\mathbf{1}]=4$ in the present matrix dimensionality. Building up infinitesimal changes in the Wess-Zumino action slowly by choosing $\alpha \rightarrow \alpha + \delta \alpha$ with $d \alpha\sim 0$ at each step, and taking contributions to first order in $\delta \alpha$ only, we allow $\alpha$ to evolve to a full profile $\theta(x,t)$ and the full Wess-Zumino action is given by
\begin{eqnarray}
S_{WZ}&=&\frac{1}{\pi}\int\theta_e (x,t) dA_{e}\\
&&+\frac{1}{\pi}\int\theta_s (x,t) dA_{S_{z}} \nonumber
\end{eqnarray}
where the charge, $\theta_e$, and spin, $\theta_s$, `axion' fields read
\begin{eqnarray}
\theta_e (x,t) &=&  \int_{0}^{\theta\left(x,t\right)} tr\left[U_{\pi/2}U^{-1}\right] \delta\alpha  \label{eq:thetae}\\
\theta_s (x,t) &=&  \int_{0}^{\theta\left(x,t\right)} tr\left[t_{3}U_{\pi/2}U^{-1}\right] \delta\alpha. \label{eq:thetas}
\end{eqnarray}

\begin{figure}[t!]
\begin{centering}
\includegraphics[scale=0.5]{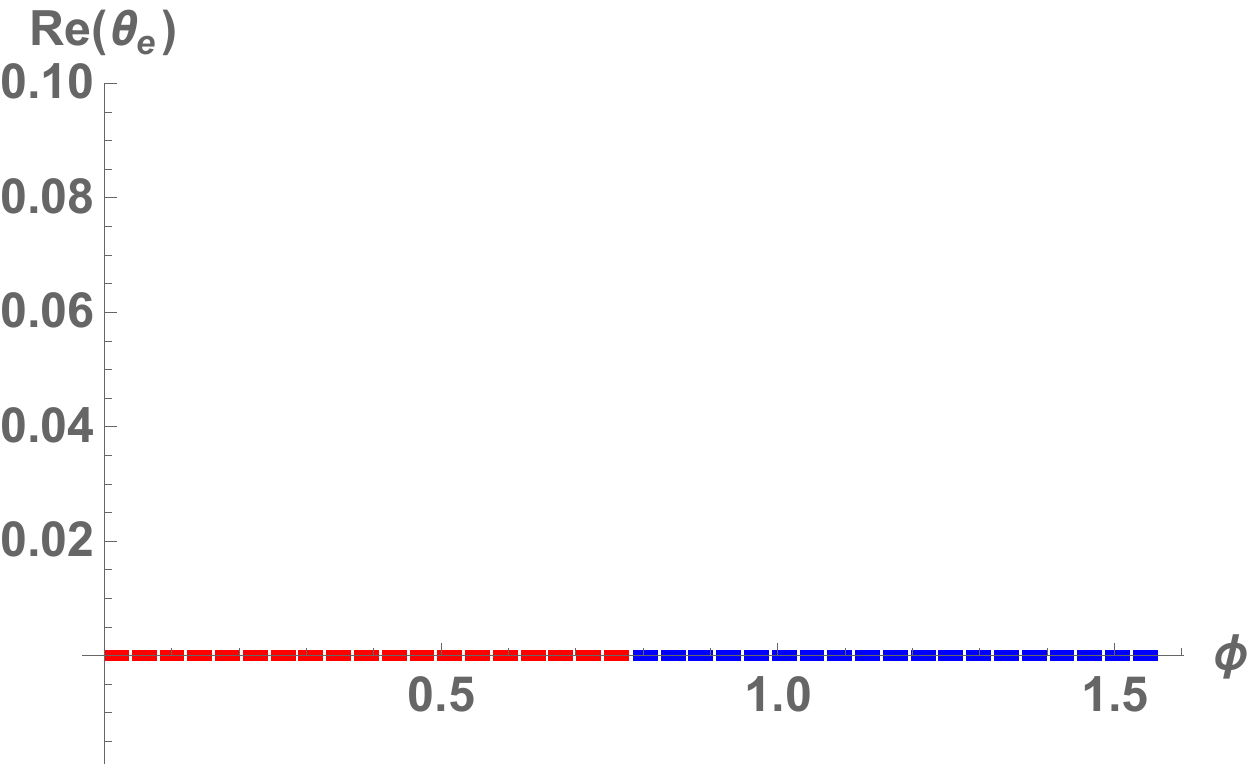}
\includegraphics[scale=0.5]{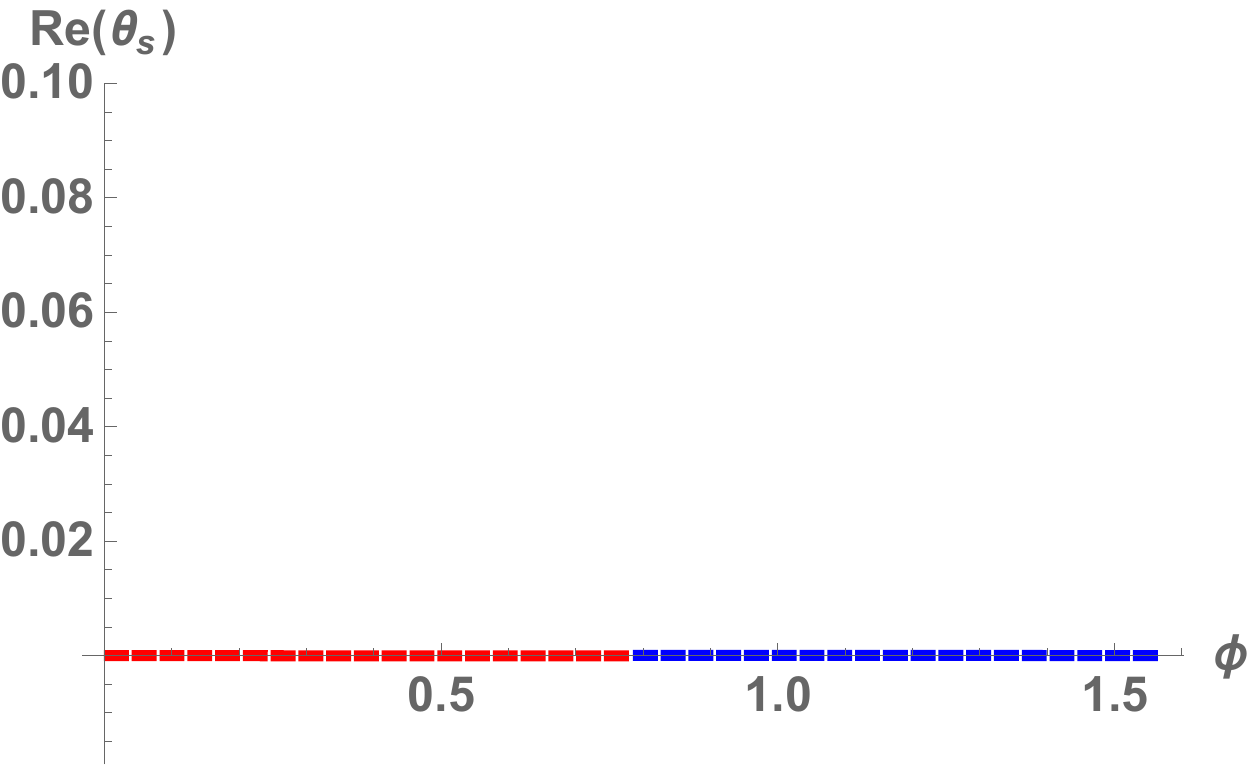}
\includegraphics[scale=0.5]{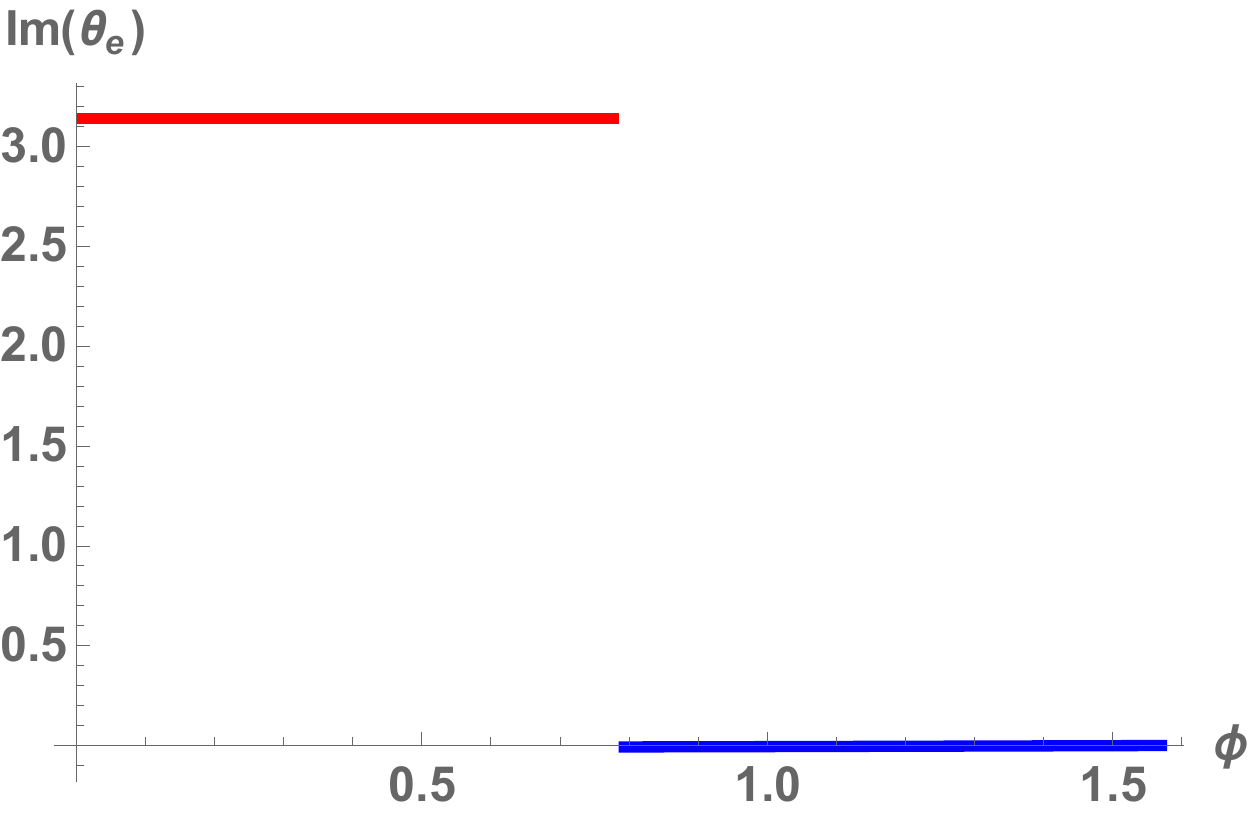}
\includegraphics[scale=0.5]{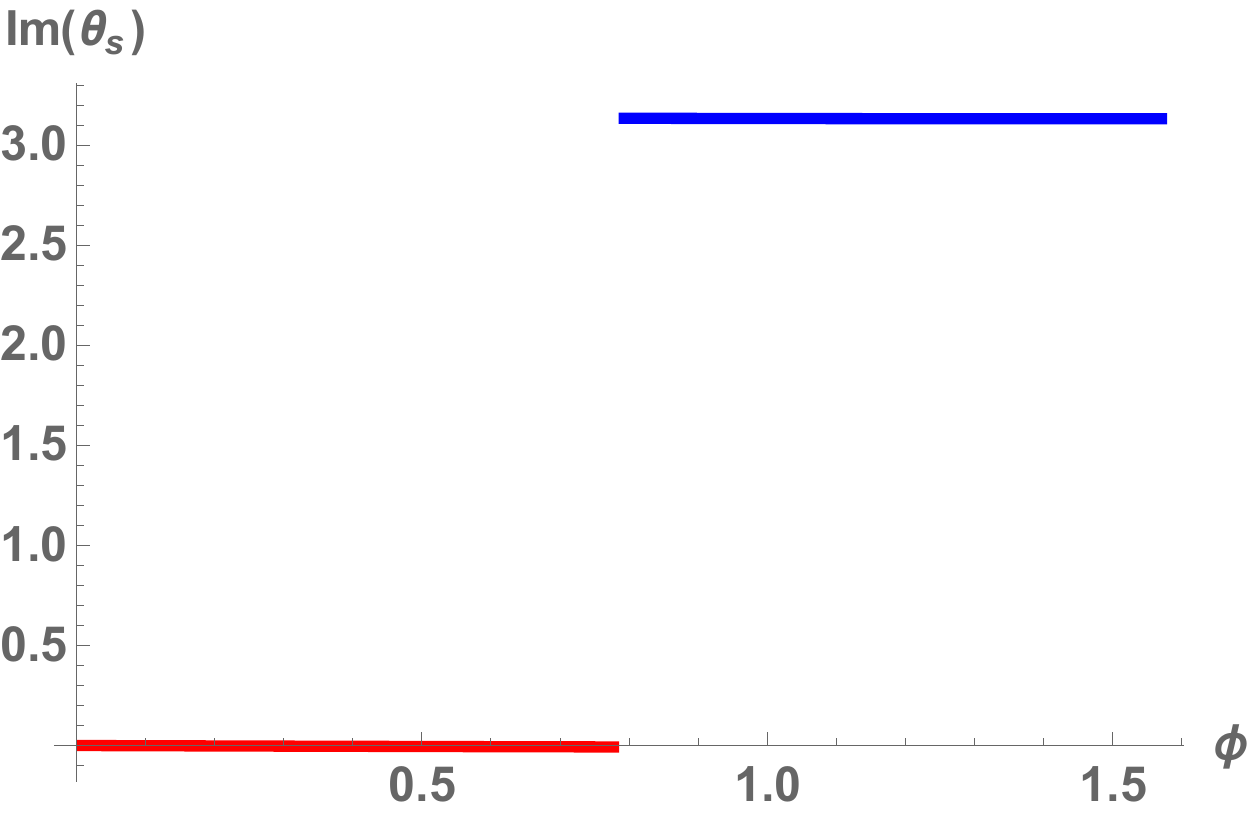}
\par\end{centering}

\protect\caption{Real and imaginary parts of $\theta_{e}$ and
$\theta_{s}$ as function of $\phi_{1}$. Notice that the real parts vanish identically while
the imaginary parts jump exactly at the gapless point with $\phi_{1}=\pi/4$.
The jump value is $\pi$. The jumping behavior for the charge and spin functions are opposite, as expected. Red and blue correspond to the topologically equivalent families of deformation paths of path I and II in Fig. \ref{fig:Hamilt_space}, respectively}
\label{fig:Ffunc1}
\end{figure}

To show that we obtain the expected results we plot in Fig.\ \ref{fig:Ffunc1} the values corresponding to full evolutions $m\rightarrow -m$ by using a fully uniform $\theta(x,t)=\pi$. We see that as a function of $\phi$, the pre-factors jump exactly by $\pi$ when $\phi$ crosses the singular line value at $\pi/4$. For path I, one sees the build up of electric charge; for path II, spin is pumped, as already discussed.

The arguments of this section generalize for the 3+1D case below, but, first, we will consider a final analysis of the 1+1D system via bosonization.

\subsection{Adiabatic Pumping Using Bosonization \label{sub:1dboson}}

Since we are considering 1+1D systems, this allows for a non-perturbative verification of the previous results using a bosonization picture of the continuum model above \cite{FradkinBook}. Considering (\ref{eq:doubSSH}), we
may proceed with the standard bosonization recipe, $\psi_{R\sigma}\sim e^{i2\sqrt{\pi}\phi_{R\sigma}},\,\psi_{L\sigma}\sim e^{-i2\sqrt{\pi}\phi_{L\sigma}}$ for bosonic fields $\phi_{R,L}.$ Generically, Klein factors would appear when
analyzing the mass terms $\Lambda_{x}$ or $\Lambda_{y},$ but to simplify our discussion we will only keep $\Lambda_z$ in this section, as to avoid discussing the non-Abelian bosonization of the full SU(2) symmetric system.  We can define the charge and spin boson fields:
\begin{eqnarray}
\phi_{c} & = & \frac{1}{\sqrt{2}}\left(\phi_{\uparrow}+\phi_{\downarrow}\right)\nonumber \\
\phi_{s} & = & \frac{1}{\sqrt{2}}\left(\phi_{\uparrow}-\phi_{\downarrow}\right).\label{eq:chargespinbos}
\end{eqnarray}
With a normalization of the boson fields by $\sqrt{2\pi}$, the Hamiltonian
(\ref{eq:doubSSH}), with $\Delta_{x,y}=0$, is mapped into the Lagrangian
\begin{eqnarray}
\mathcal{L} & = & \frac{1}{4\pi}\left(\partial_{\mu}\phi_{c}\right)^{2}+\frac{1}{4\pi}\left(\partial_{\mu}\phi_{s}\right)^{2} \nonumber\\
 &  & -\frac{e}{\pi}\phi_{c}\epsilon^{\mu\nu}\partial_{\nu}A_{\mu} -\frac{1}{2\pi}\phi_{s}\epsilon^{\mu\nu}\partial_{\nu}A_{\mu}^{S_z}\nonumber \\
 &  & -m\frac{2}{\pi a}\cos\left(\phi_{c}\right)\cos\left(\phi_{s}\right)\nonumber \\
 &  & +m_{5}\frac{2}{\pi a}\sin\left(\phi_{c}\right)\cos\left(\phi_{s}\right)\label{eq:BosonLag}\\
 &  & +\Delta_{z}\frac{2}{\pi a}\cos\left(\phi_{c}\right)\sin\left(\phi_{s}\right),\nonumber
\end{eqnarray}
where  $A^{\mu}=\left(A^{0},\, A^{1}\right)$ is the electromagnetic vector potential, $A_{\mu}^{S_z}$ is the corresponding spin-gauge field, and
$a$ is a short-distance cut-off scale.

One of the advantages of the bosonized
picture is that we can see explicitly how the electromagnetic response
of the system is associated only with a charge-like bosonic field $\phi_{c},$
 and that there is no coupling between $\phi_{s}$ and $A^{\mu}$. Hence, we would not expect that shifts of $\phi_s$ would generate any electromagnetic response, something that is not as obvious in the perturbative field theory calculations of the Dirac fermions above. A similar analysis holds or the spin-gauge field response.

Since the ground state configurations in the bosonized picture are simple, this allows us to
study the properties of the full parameter space, even where perturbative calculations of the Dirac fermions would break down close
to the gapless singular lines in the adiabatic parameter space. At low-energies, a semi-classical study of the problem is enough for our purposes. First, the kinetic terms guarantee that
the ground states should be homogeneous constant fields, $\phi_{c}=const.$
and $\phi_{s}=const.^{'}$. Now, let us first consider $m_{5}=\Delta_{z}=0$.
One reads from the potential
\begin{equation}
\sim m\cos\left(\phi_{c}\right)\cos\left(\phi_{s}\right)
\end{equation}
that there are two degenerate minima for each sign of $m$, namely
\begin{eqnarray}
\left(\phi_c,\phi_s\right)&=&\left\{\begin{array}{ccc} (0,\pi)\\ {\rm{or}} \\(\pi,0)\end{array}\right. {\rm{if}}\; m>0\nonumber\\
\left(\phi_c,\phi_s\right)&=&\left\{\begin{array}{ccc} (0,0)\\ {\rm{or}}\\(\pi,\pi)\end{array}\right. {\rm{if}}\; m<0.
\end{eqnarray}
 As one might expect, $m=0$ leaves an ill-defined (gapless) ground
state. We note that to adiabatically connect the $m>0$ pair to the $m<0$ pair we need to
consider a path with finite $m_{5}$ or $\Delta_{z}$ in the Hamiltonian parameter
space.

A possible question which may arise at this point is: why does the bosonized version of the closed/periodic (1+1)D problem have two ground states (for a given mass sign) while the original free fermion problem only appears to have one? The answer is that in our original fermionic problem we did not discuss the possibilities of different types of periodic boundary conditions for the fermions (i.e., different spin structures). Indeed, distinct pinned bosonic ground states can be accessed by piercing the closed fermionic chain by $\pi-$fluxes of the charge/spin gauge fields. Such fluxes flip the boundary conditions from periodic to anti-periodic, hence flipping the parity of the fermionic ground states. For our system there are two parity operators, giving four possibilities for the ground states. These are the states manifest in the present bosonic language.

\begin{figure}[t]
\subfloat[]{\begin{centering}
\includegraphics[scale=0.25]{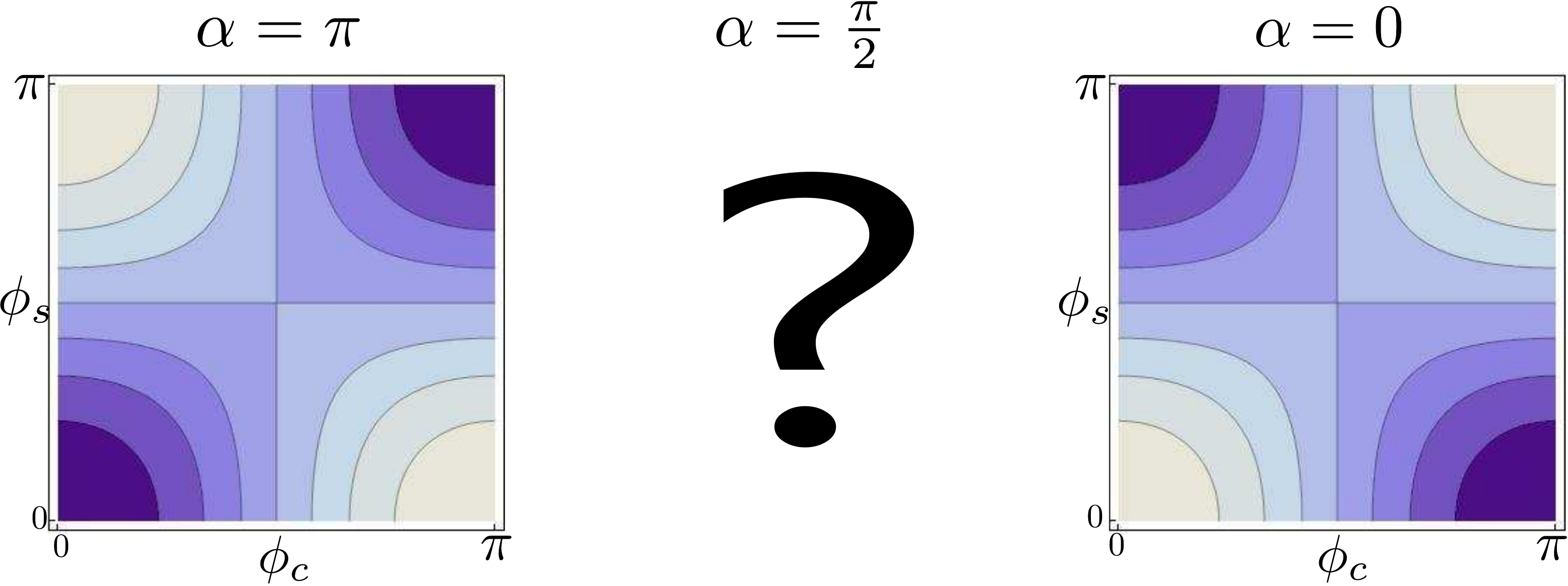}
\par\end{centering}

}

\subfloat[]{\begin{centering}
\includegraphics[scale=0.25]{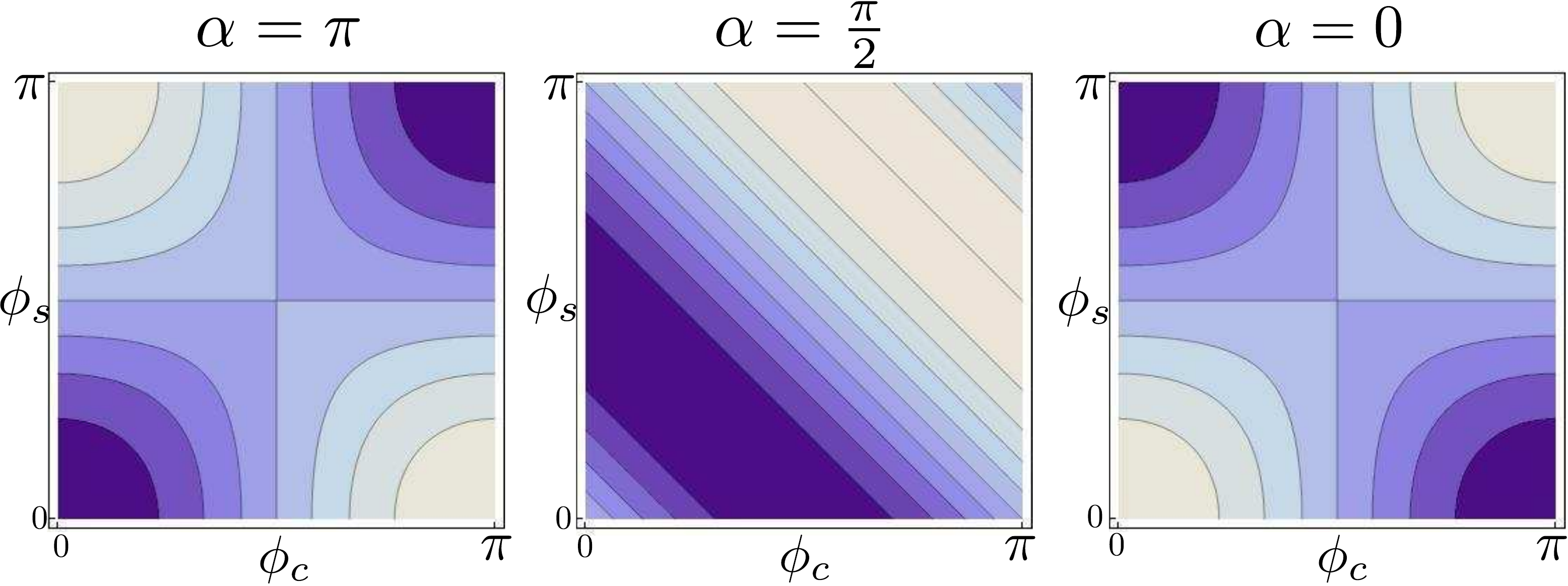}
\par\end{centering}

}

\caption{Evolution of the ground states for $m=1$. Symmetry allows us to consider
only values of $0\leq\phi_{c},\,\phi_{s}\leq\pi$. The dark blue circles
correspond to the minima of $V\left(\phi_{c},\phi_{s},\alpha\right)$
for $m>0$ while the white circles correspond to the minima for $m<0$.
(a) $m_{5}=\Delta_{z}=0$; at $\alpha=\pi/2$ it is impossible to
define the ground state values for $\phi_{c}$ and $\phi_{s}$. (b)
$m_{5}=\Delta_{z}=1$; at $\alpha=\pi/2$ the minima become connected
in phase space.\label{fig:GroundStatesIll}}
\end{figure}

Now let us consider adiabatic evolution. We can nicely parameterize our paths of interest by the potential
\begin{eqnarray}
V\left(\phi_{c},\phi_{s},\alpha\right) & = & m\cos\alpha\cos\left(\phi_{c}\right)\cos\left(\phi_{s}\right)\nonumber \\
 &  & -m_{5}\sin\alpha\sin\left(\phi_{c}\right)\cos\left(\phi_{s}\right)\nonumber \\
 &  & -\Delta_{z}\sin\alpha\cos\left(\phi_{c}\right)\sin\left(\phi_{s}\right).
\end{eqnarray}
With $\alpha$ evolving from $0$ to $\pi$, we may consider the two different sets of $\pi$-rotations
from Section \ref{sub:1DGeneralities}. If $\Delta_{z}=0$, this evolution
corresponds to the usual chiral rotation which, in the bosonization
formalism, is translated to $\phi_{\sigma}\rightarrow\phi_{\sigma}+\frac{\pi}{\sqrt{2}}.$
Then, from (\ref{eq:chargespinbos}),
\begin{eqnarray}
\phi_{c} & \rightarrow & \phi_{c}+\pi\nonumber \\
\phi_{s} & \rightarrow & \phi_{s},
\end{eqnarray}
and we see that this transformation connects the ground states
 $(0, \pi)\to(\pi, \pi)$ or the ground states $(\pi,  0)\to (
0, 0).$ In either case, the change in the Lagrangian is simply
\[
\left|\Delta\mathcal{L}\right|=e\epsilon^{\mu\nu}\partial_{\nu}A_{\mu}
\]
which, as expected, corresponds to twice the 1D $\theta$-term, i.e., the same result found in Eq. \ref{eq:1DpathI}. Hence, rotating $\phi_c$ by $2\pi$ recovers the adiabatic charge pumping mechanism that was discussed above in a fermion language.

Analogously, if $m_{5}=0$ and one follows the evolution in the $m-\Delta_z$-plane, one finds
\begin{eqnarray}
\phi_{c} & \rightarrow & \phi_{c}\nonumber \\
\phi_{s} & \rightarrow & \phi_{s}+\pi.
\end{eqnarray}
The ground states are now connected as  $(0, 0)\to(\pi, \pi)$ or $(\pi,  0)\to (
0, \pi),$  and there is
no change in $\phi_{c}$. In fact, the additionally imposed $U(1)$ \emph{spin} gauge field coupled to the conserved spin component  generates a $\theta$-term for the spin gauge field, and shifting $\phi_s$ by $2\pi$ acts as an adiabatic spin pump.

\begin{figure}[t]
\subfloat[]{

\begin{centering}
\includegraphics[scale=0.25]{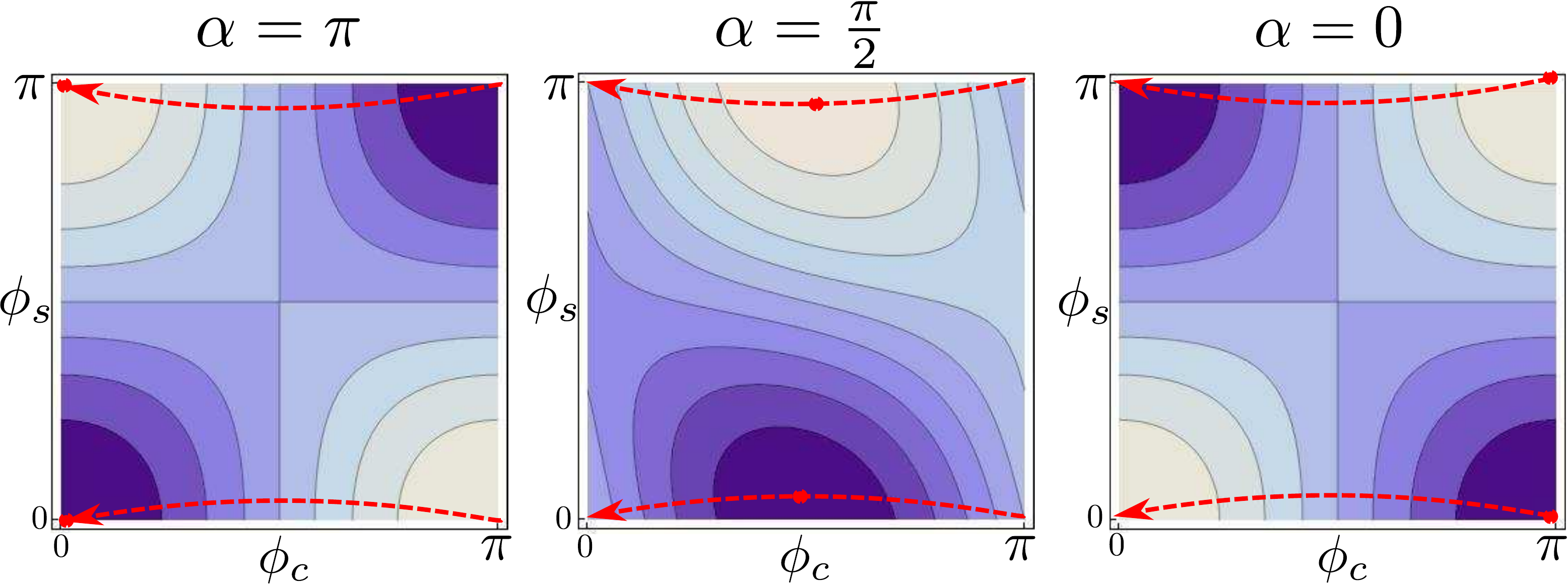}
\par\end{centering}

}

\subfloat[]{

\begin{centering}
\includegraphics[scale=0.25]{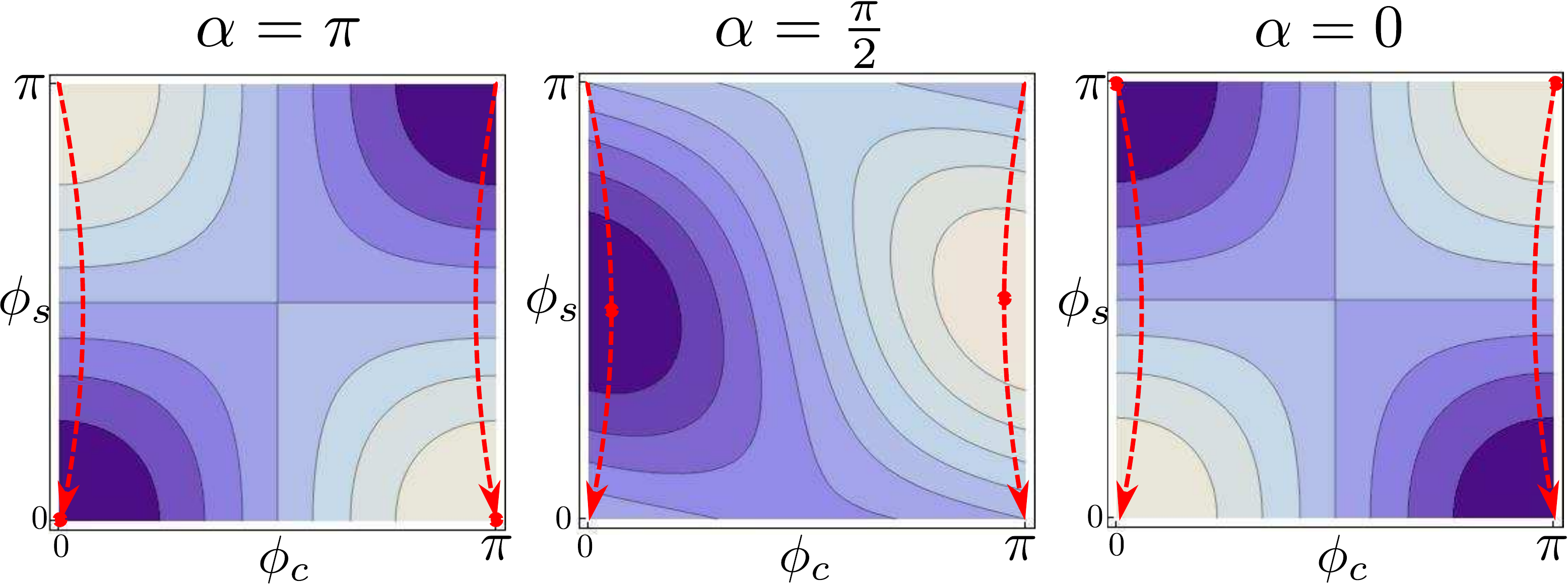}
\par\end{centering}

}\caption{Evolution of the ground states for $m=1$. Symmetry allows us to consider
only values of $0\leq\phi_{c},\,\phi_{s}\leq\pi$. The dark blue circles
correspond to the minima of $V\left(\phi_{c},\phi_{s},\alpha\right)$
for $m>0$ while the white circles correspond to the minima for $m<0$.
The red dashed lines show the trajectories of the minima, red dots,
as $\alpha$ is swept from $0$ to $\pi$ (a) $m_{5}=1$, $\Delta_{z}=0.25$,
during the evolution the minima move mostly in the horizontals but
connects the two pairs of ground states uniquely. (b) $\Delta_{z}=1$,
$m_{5}=0.25$, In this case, the minima move mostly in vertical but
again the ground states are always well defined.\label{fig:GroundStatesWell}}

\end{figure}

Now, if one considers finite values of both $m_{5}$ and $\Delta_{z}$,
as we evolve $\alpha$, the minima in $(\phi_c,\phi_s)$-space will start
to trace more complex trajectories,  straying away from
the horizontal and vertical lines of the two special cases above. However, as long as $\vert m_5\vert \neq \vert \Delta_z \vert$ then the starting and ending points match the results for ${\rm{max}}(\vert m_5\vert, \vert \Delta_z\vert).$ Also, despite the degeneracy of the ground state for a fixed sign of $m$, mixing between the
ground states never happens. On the other hand, if  $\vert m_{5} \vert=\vert \Delta_{z}\vert$ then
during the path, when $\alpha=\pi/2$, we find a valley of degenerate
minima connecting the original two minima of the system, thus allowing
for a non-adiabatic change in the ground state, and making the evolution ill defined,
as expected. These considerations are summarized in Figs. \ref{fig:GroundStatesIll}
and \ref{fig:GroundStatesWell}.

Interestingly, we can also consider the hybrid path discussed earlier where we first rotate in the $m-m_5$-plane by $\pi$ and then in the $m-\Delta_z$-plane by $-\pi$ to return to the starting point. In this case we see that during this process we have the transformation
\begin{eqnarray}
\phi_{c} & \rightarrow & \phi_{c}+\pi\nonumber \\
\phi_{s} & \rightarrow & \phi_{s}+\pi.
\end{eqnarray} Hence, for a fixed sign of $m,$ this process keeps us in the same degenerate ground state subspace but switches among the two ground states. Thus, we can see that the singular line in our parameter space acts as a kind of non-Abelian Berry flux that acts to switch the ground state. Encircling the singular line twice will return us to the original ground state.

\section{ 3+1D Doubled TI\label{sec:3+1D}}

So far we have studied generalized adiabatic pumping processes in doubled 1D systems. For this, we computed electromagnetic- and spin-gauge effective responses for 1D insulating systems via adiabatic transformations of the Hamiltonian. From the doubled model we were able to develop inequivalent classes of  paths in parameter space. Depending on the path chosen, the effective action implied that the adiabatic transformation led to a charge-pump, a spin-pump, or a ``full-electron" pump.

The general results and procedures in 1D and 3D are very similar. From the point of view of electromagnetic responses, adiabatic pumping in 3D is signaled by the Chern-Simons axion coupling
\begin{equation}
S_{eff}[A]=\frac{1}{8\pi^{2}}\int\theta dAdA,
\end{equation}
where $A$ is the electromagnetic vector potential, and $\theta$ parametrizes the adiabatic deformation of the Hamiltonian. The corresponding currents derived from this effective action are
\begin{equation}
\langle j^{\mu}\rangle = \frac{\delta S_{eff}}{\delta A_\mu}=\frac{1}{4\pi^2}\epsilon^{\mu\nu\rho \sigma }\partial_\nu \theta(x,t)\partial_\rho A_\sigma.
\end{equation}
To generate a response one needs both a space or time variation of the adiabatic parameter,  and an electric or magnetic field, respectively. The physics of this axion coupling is exactly the topological magneto-electric effect that exists in 3D time-reversal invariant topological insulators\cite{QHZ2008}. In this context, the corresponding pumping process obtained by the evolution of the $\theta$ parameter has been dubbed a "Chern-Simons axion adiabatic pumping" \cite{Taherinejad}.

An simple way to understand this response can be developed in the presence of a uniform magnetic field. In this case, the response becomes analogous to copies of the response for a 1D $Z_2$ topological insulator, i.e., the usual Goldstone-Wilczek response. The number of copies is simply equal to the number of flux quanta.  We will adopt this idea and use it to explore the different quantum numbers pumped in this context.

\subsection{Generalities \label{sub:3dgeneral}}

Consider doubling a minimal model for a 3D TI Hamiltonian and including additional mass terms. For example, we study the model
\begin{eqnarray}
\mathcal{H}_{3D} & = & \int  d^3x\, \Psi^{\dagger}H_{3D}\Psi\nonumber \\
H_{3D} & = & \boldsymbol{\Gamma}\cdot\left(\mathbf{p}-\mathbf{A}\right)+\Gamma_{0}m+\Lambda_{5}m_{5}+\boldsymbol{\Lambda}\cdot\boldsymbol{\Delta},\label{eq:3DTI_SC}
\end{eqnarray}
now with the basis
\begin{eqnarray}
\Gamma_{1}=\rho_{0}\alpha^{x},\quad  &  & \,\, \Lambda_{5}=\rho_{0}\beta_{5}\nonumber \\
\Gamma_{2}=\rho_{0}\alpha^{y},\quad  &  & \,\, \Lambda_{x}=\rho_{x}\beta_{5}\nonumber \\
\Gamma_{3}=\rho_{0}\alpha^{z},\quad  &  & \,\, \Lambda_{y}=\rho_{y}\beta_{5} \label{eq:3dmatrices}\\
\Gamma_{0}=\rho_{0}\beta,\quad  &  & \,\, \Lambda_{z}=\rho_{z}\beta_{5}\nonumber
\end{eqnarray}
where
\begin{eqnarray}
\boldsymbol{\alpha} & = & \tau_{z}\boldsymbol{\sigma}\nonumber \\
\beta & = & \tau_{x}\sigma_{0}\\
\beta_{5} & = & \tau_{y}\sigma_{0}.\nonumber
\end{eqnarray}
In this basis, $\sigma_{i}$ represents spin, $\tau_{i}$ are orbital
degrees of freedom, and $\rho_{i}$ are the doubling-related degrees of freedom (which we do not specify explicitly, but may be of orbital content also, and will henceforth be referred to as ``doubling-orbital" degrees of freedom).
This model has discrete time-reversal and particle-hole symmetries with operators $\mathcal{T}=i\sigma_{y}K$ and  $\mathcal{C}=\rho_{y}\tau_{y}\sigma_{y}K$.  
Table \ref{tab:3DSymmetries} summarizes the symmetry properties of the extra mass terms with respect to these operations. These additional mass terms will define the pumping parameter space as discussed below. As an aside, if one were to identify $\mathcal{H}_{3D}$ with a BdG doubled Hamiltonian, such that $\mathcal{C}$ were a strict particle-hole redundancy, then this model is just a 3D topological insulator with a chiral mass $m_5$, and $\Delta_{y}$ and $\Delta_{x}$ are the real and imaginary parts of a proximity-coupled s-wave superconducting order parameter. As the electromagnetic charge U(1) symmetry is broken in this case, the concept of pumping is not as well defined and we leave further comments on this to the conclusion. Another interpretation for this model is that of a topological crystalline insulator. This system has a trivial strong invariant, but will have non-vanishing mirror Chern numbers for the mirror operators given by $\hat{m}_x=i\Gamma_{x}\Lambda_5, \hat{m}_y=i\Gamma_{y}\Lambda_5,$ and  $\hat{m}_z=i\Gamma_{z}\Lambda_5$. We can see this by restricting the model to the $k_x=0$, $k_y=0$ or $k_z=0$ planes respectively and calculating the mirror Chern number which we find to be $C_{x}=2$, $C_{y}=2,$ or $C_{z}=2$ respectively, if we keep our previous choice of regularization where a negative mass represents the topological phase.  This implies that on surfaces with $\hat{m}_x$, $\hat{m}_y$ or $\hat{m}_z$ symmetry there will be two surface Dirac cones stabilized by mirror symmetry that lie along a mirror line; recall that only an odd number of cones is stabilized if we just have time-reversal symmetry. 

\begin{table}[t]
\begin{centering}
\begin{tabular}{c|c|c}
\hline
Matrix & C & T\tabularnewline
\hline
{\large{$\Lambda_{5}$}} & {\large{$\checkmark$}} & {\large{$\times$}}\tabularnewline
{\large{$\Lambda_{x}$}} & {\large{$\checkmark$}} & {\large{$\times$}}\tabularnewline
{\large{$\Lambda_{y}$}} & {\large{$\checkmark$}} & {\large{$\checkmark$}}\tabularnewline
{\large{$\Lambda_{z}$}} & {\large{$\times$}} & {\large{$\times$}}\tabularnewline
\hline
\end{tabular}
\par\end{centering}

\caption{Symmetry properties of the masses for the (3+1)D doubled TI model under charge-conjugation $\left(C\right)$
and time-reversal $\left(T\right)$ transformations. The corresponding symmetry is broken/preserved for $\times$/$\checkmark$ marks, respectively. Our choice of symmetry structure leads to an exactly mirrored behavior between the C and T columns with respect to the 1D case. \label{tab:3DSymmetries}}
\end{table}

When exploring the response and pumping processes, many of the results from Sec.\ \ref{sub:1DGeneralities} hold in the 3D case with few modifications. The extra mass terms are again compatible with the usual TI mass, but compete with each other. We find that, of example, the evolution through path $I$ computed from the standard triangle anomaly using the $m_5$ mass gives twice the standard $\theta$-term, i.e.,
\begin{equation}
S_{eff}^I= 2\Theta_{3D},
\end{equation}
where
\begin{equation}
\Theta_{3D}=\frac{e^{2}}{4\pi^{2}}\int d^4x \, \theta\mathbf{E}\cdot\mathbf{B}.\label{eq:3Dtheta}
\end{equation}
All of the responses can be calculated similarly to our method in 1D, and we explicitly show this in Appendix \ref{app:Appendix-A}. Table \ref{fig:-Summary3DBdG} summarizes the results obtained for the doubled 3D case using these perturbative methods. In what follows, we give a microscopic interpretation for the (3+1)D case.

\begin{figure}
\begin{centering}
\begin{tabular}{|c|c|}
\hline
Path & Response\tabularnewline
\hline
$\mathrm{I}$ & $2 \Theta_{3D}$\tabularnewline
\hline
$\mathrm{II}$ & $-$\tabularnewline
\hline
$\mathrm{III}$ & $2 \Theta_{3D}$\tabularnewline
\hline
$\mathrm{IV}$ & $-$\tabularnewline
\hline
\end{tabular}
\par\end{centering}

\caption{ Summary of the responses calculated in each path of Fig. \ref{fig:Hamilt_space}
for the particle-hole doubled (3+1)D case. \label{fig:-Summary3DBdG}}
\end{figure}

\subsection{3D TI surfaces in a magnetic field \label{sub:3dsurf}}

In this section we aim to connect the results found in (3+1)D in Sec.\ \ref{sub:3dgeneral}
with those in Sec.\ \ref{sec:1+1D} for the (1+1)D system, so that we can easily interpret
the types of quantum numbers are pumped on each class of paths in our adiabatic parameter space.  We will consider this in the context of the doubled 3D TI response (\ref{eq:3Dtheta})
in the case of a uniform magnetic field, say, in the $z$ direction,
\begin{equation}
\Theta_{3D}=\frac{e^{2}B_{z}}{4\pi^{2}}\int d^4x \theta E_{z}.
\end{equation}
If we define $\bar{E}_{z}=\int dxdyE_{z}/A_{xy}$ to be the average of
the electric field in the $xy$ plane with area $A_{xy}$, we find
\begin{equation}
\Theta_{3D}=\frac{e}{2\pi}\int dzdt\tilde{\theta}\bar{E}_{z},
\end{equation}
with $\tilde{\theta}=\frac{e\Phi}{2\pi}\theta=N_{\Phi}\theta,$  $\Phi=B_{z}A_{xy}$ is the magnetic flux through the plane, and $N_{\Phi}$ is the number of flux quanta in units of $\Phi_0=2\pi/e (=h/e).$ Because of this form of the effective action we expect that the surfaces of the doubled 3D TI in the presence of a uniform magnetic field must behave analogously to ($N_\Phi$ copies of) the ends of the doubled SSH model. In particular, the spin
and charge pumps considered in Sec.\ \ref{sec:1+1D} should have
an analogy in (3+1)D.

To see this we can solve the surface Dirac Hamiltonian explicitly when in the presence of the uniform perpendicular magnetic field. The surface physics of the 3D TI is known for being described by massless
Dirac fermions,  and the surface modes may be found by considering a TI mass with a domain
wall as a simple model for the surface
\begin{equation}
m\left(z\right)=\begin{cases}
<0 & if\ z>0\\
>0 & if\ z<0
\end{cases},
\end{equation}
and looking for states localized around $z=0$. An ansatz for these localized states takes the form
\begin{equation}
\psi\left(\mathbf{r}\right)=\frac{1}{\sqrt{\mathcal{N}}}e^{-\int_{0}^{z}m(z^{\prime})dz^{\prime}}\phi\left(x,y\right),
\end{equation}
where $\mathcal{N}$ normalizes the exponential factor, and $\phi$ is a normalized spinor satisfying
\begin{equation}
m\left(z\right)\left(-i\rho_{0}\alpha_{z}+\rho_{0}\beta\right)\phi=0.
\end{equation} For the doubled system,
the spinors are of the form $\phi=\left( \phi_{1} , \phi_{2} \right)^T$, where $\phi_{1,2}$ are four-component spinor wavefunctions of only the $x,y$ coordinates. This reduces the matrix dimension of the problem by two, whose solution allows for the definition
of a basis for the localized states
\begin{equation}
\chi_{1}=\frac{1}{\sqrt{2}}\left(\begin{array}{c}
1\\
0\\
i\\
0
\end{array}\right),\ \chi_{2}=\frac{1}{\sqrt{2}}\left(\begin{array}{c}
0\\
1\\
0\\
-i
\end{array}\right),\label{eq:surfsubbasis}
\end{equation}
and the basis
\begin{equation}
\left(\begin{array}{c}
\chi_{1}\\
0
\end{array}\right),\,\left(\begin{array}{c}
\chi_{2}\\
0
\end{array}\right),\,\left(\begin{array}{c}
0\\
\chi_{1}^{*}
\end{array}\right),\,\left(\begin{array}{c}
0\\
\chi_{2}^{*}
\end{array}\right).\label{eq:surfbasis}
\end{equation}
Now we may project the original Hamiltonian \eqref{eq:3DTI_SC} in the
new basis, to obtain the surface Hamiltonian
\begin{equation}
\tilde{H}_{surf}=\tilde{\boldsymbol{\Gamma}}\cdot\mathbf{p}_{\perp}-\tilde{\boldsymbol{\Gamma}}\cdot \tilde{\mathbf{A}}_{\perp}+\tilde{\Lambda}_{5}m_{5}+\tilde{\boldsymbol{\Lambda}}\cdot\boldsymbol{\Delta},\label{eq:surfH}
\end{equation}
where $\perp$ fixes only components in the $xy$ plane. The projected matrices are
\begin{eqnarray}
\tilde{\Gamma}^{x} & = & \rho_{0}\tilde{\sigma}_{x}\nonumber \\
\tilde{\Gamma}^{y} & = & \rho_{0}\tilde{\sigma}_{y}\nonumber \\
\tilde{\Lambda}_{5} & = & \rho_{0}\tilde{\sigma}_{z}\nonumber \\
\tilde{\Lambda}_{1} & = & \rho_{x}\tilde{\sigma}_{y}\nonumber \\
\tilde{\Lambda}_{2} & = & \rho_{y}\tilde{\sigma}_{y}\nonumber \\
\tilde{\Lambda}_{3} & = & \rho_{z}\tilde{\sigma}_{z},
\end{eqnarray}
where $\tilde{\sigma}$ and $\rho$ are 2$\times$2 matrices
in the basis (\ref{eq:surfbasis}). Notice that $\rho$ still
represents the doubling degrees of freedom of the original Hilbert space, while, from (\ref{eq:surfsubbasis}),
the states $\chi_{1,2}$ have well defined spin (up/down respectively). The gauge field, being z-independent, is unaffected by $z$-averaging
\begin{equation}
\tilde{\mathbf{A}}_{\perp} \equiv \int_{-\infty}^{+\infty}\frac{1}{\mathcal{N}} e^{-2\int_{0}^{z}m\left(z^{'}\right)dz^{'}}  \mathbf{A}_{\perp} =\mathbf{A}_{\perp}.
\end{equation}
 
Now we may study the Landau level problem of the surface, our aim being to find the low-energy degrees of freedom bound to the surfaces in a magnetic field, and then to consider the effects of masses $m_5,$ and $\boldsymbol{\Delta}$ perturbatively as we did for the end states of the (1+1)D system. For the surface, the blocks of the doubled degrees of freedom decouple and one has to solve the eigenvalue problem for
\begin{equation}
H_{QHE,i}=\tilde{\sigma}_{x}\left(p_{x}-A_{x}\right)+\tilde{\sigma}_{y}\left(p_{y}-A_{y}\right),
\end{equation}
with $i=1,2$ fixing the two surface copies, and where we have replaced $\tilde{\mathbf{A}}$ with ${\bf{A}}$ since they are equivalent here.
The solution to this problem is known and is analogous to studying the Landau level problem in graphene\cite{Neto2009}. Diagonalization may be achieved by considering $\mathbf{A}$
in the symmetric gauge $\mathbf{A}=-\frac{\mathbf{r}\times\mathbf{B}}{2}=\frac{B}{2}\left(-y, \, x, \,0 \right)$ and writing complex coordinates $\xi=x+iy$. With these one defines
the ladder operators
\begin{equation}
O_{i}^{\dagger}=\frac{1}{\sqrt{2}}\left(-i\partial_{\xi}+i\xi^{*}\right),\ O_{i}=\frac{1}{\sqrt{2}}\left(-i\partial_{\xi^{*}}-i\xi\right)
\end{equation}
acting on each copy of the surface, such that
\begin{eqnarray}
H_{QHE,i}\left|\chi_{i}\right\rangle =\left(O_{i}^{\dagger}\tilde{\sigma}^{+}+O_{i}\tilde{\sigma}^{-}\right)\left|\chi_{i}\right\rangle  & = & \frac{E}{\omega_{c}}\left|\chi_{i}\right\rangle,
\end{eqnarray}
with $\tilde{\sigma}^\pm\equiv \frac{1}{2}(\tilde{\sigma_{x}}\pm i\tilde{\sigma_{y}})$ and the cyclotron frequency in the present units reads $\omega_c=\sqrt{2B}.$ 

Squaring the Hamiltonian shows that the number operators $O_{i}^{\dagger}O_{i} $ are good
quantum numbers and one can solve the Hamiltonian to find the energies to be $E_{\pm N}=\pm\omega_{c}\sqrt{N}$.
Of importance for our analysis are just the, lowest energy, zero modes
\begin{equation}
\tilde{\chi}_{i,0}=\left(\begin{array}{c}
1\\
0
\end{array}\right)\left|0\right\rangle_i, \label{eq:chip}
\end{equation}
where $\left\langle \xi,\xi^{*}|0\right\rangle_i$ is the zero-mode wave function in the $i$-th copy. Notice that in our surface of choice, the $\tilde{\chi}_{i,0}$ spinor is totally polarized in the upper $\chi_1$  space, which we see from the previous analysis, carries spin $\uparrow$. Hence, both copies of the surface state zero-modes are totally up-spin polarized; they will be down polarized on the other surface.

Now we can treat the mass terms perturbatively in the subspace of degenerate zero modes, including each copy. The projection into the zero-mode subspace implies
\begin{eqnarray}
\tilde{\Gamma}^{x} & \rightarrow & 0\nonumber \\
\tilde{\Gamma}^{y} & \rightarrow & 0\nonumber \\
\tilde{\Lambda}_{5} & \rightarrow & \rho_{0}\nonumber \\
\tilde{\Lambda}_{1} & \rightarrow & 0\nonumber \\
\tilde{\Lambda}_{2} & \rightarrow & 0\nonumber \\
\tilde{\Lambda}_{3} & \rightarrow & \rho_{z}.
\end{eqnarray}
Assuming that the surface we considered was the upper one, we write the low-energy surface Hamiltonian
\begin{equation}
\bar{H}_{u}=m_{5} \rho_{0}+\Delta_z \rho_{z},\label{eq:lowensurfH}
\end{equation}
where $u$ stands for the upper surface. To closely compare with our (1+1)D lattice model results
we need to also consider the lower surface. The projection for the lower surface can be found in practice by simply inverting the sign of the external magnetic field, such that spin will then point anti-parallel to the surface. We may finally write
\begin{equation}
H_{surf}=\left(\begin{array}{cccc}
m_{5}+\Delta_{z} & 0 & 0 & 0\\
0 & m_{5}-\Delta_{z} & 0 & 0\\
0 & 0 & -m_{5}-\Delta_{z} & 0\\
0 & 0 & 0 & -m_{5}+\Delta_{z}
\end{array}\right),\label{eq:Hedge}
\end{equation}
in the basis $\left\{ \begin{array}{cccc}
\left|u_\uparrow,1\right\rangle  & \left|u_\uparrow,2\right\rangle  & \left|l_\downarrow,1\right\rangle  & \left|l_\downarrow,2\right\rangle \end{array}\right\} $, where $u_\uparrow,\, l_\downarrow$ label the up-spin upper surface and down-spin lower surface.

With this form of the perturbative Hamiltonian we could repeat a discussion analogous to the one 
in Sec.\ \ref{sub:Microscopic1D} for the 1D case. In fact, comparing $H_{surf}$ derived in here with $H_{edge}$ from  Sec.\ \ref{sub:Microscopic1D} shows the exact same result, the only difference being the quantum numbers carried by the end modes, and the fact that there is a Landau level degeneracy in 3D equal to the number of magnetic flux quanta penetrating the surface. In the present case, the interplay between the strength of $m_{5}$ and $\Delta_z$ determines whether we will have spin polarization on the surface, or a polarization in the doubling-orbital degrees of freedom.

Going back to our interest in pumping processes, these surface effects imply that evolutions through path Ib, from Fig.\ref{fig:Berry-phase} should correspond to pumps of the (surface) electromagnetic Chern-Simons coefficient, i.e., an electromagnetic surface Hall conductivity, i.e., one complete cycle pumps a Hall conductivity of $|2e^2/h|$. As we will see in the next section, this effect is also accompanied by a pumping in the surface Chern Simons term for the gauge field coupling to the conserved $U(1)$ charge of the doubled-orbital degree of freedom. Interestingly, the corresponding effect for path Ia is that the electromagnetic field may pump a mixed Chern Simons term between the electromagnetic and doubled-orbital gauge fields. In this case, electric (magnetic) fields of one gauge field will drive a Hall current (trap charge) of the other gauge field. This suggests that the bulk action in this case should contain a mixed axion term between gauge fields corresponding to the electromagnetic field and the conserved corresponding charge from the doubled-orbital since this translates to a mixed Chern-Simons term between these gauge fields on the surface. We will see in the next Section that this is realized precisely. Finally, evolutions through paths IIa and IIb should correspond to a combination of pumps between the regular Hall (electromagnetic and doubled-orbital) coefficients and the mixed Hall ones, with opposite signs between IIa and IIb. 

\subsection{Ma\~{n}es-Bardeen Anomaly in 3D\label{sub:bardeen}}
In three spatial dimensions, the bosonization analysis is not easily available. However, it is possible
to repeat the anomaly analysis derived from the Bardeen form of the gauged Wess-Zumino action\cite{BardeenZumino, Manes, VPNair}. This will allow for a non-perturbative check on our perturbative response and pumping results, e.g., to confirm there is no axion-term build up in $\Delta$-mass type evolutions of the Hamiltonian.

The analysis of the gauge symmetries, the action, and paths follows exactly as in Sec.\ref{sub:1dmanes}, as that is dimension independent. In our representation of the 3D system the SU(2) matrices now read
\begin{eqnarray}
t_{1}=\frac{1}{2}\rho_{x}\tau_{y}\sigma_{y},\quad 
&&\lambda_{1}=-\frac{1}{2}\rho_{y}\tau_{x}\sigma_{y}\\t_{2}=\frac{1}{2}\rho_{y}\tau_{y}\sigma_{y},\quad 
&&\lambda_{2}=\frac{1}{2}\rho_{x}\tau_{x}\sigma_{y}\\t_{3}=\frac{1}{2}\rho_{z}\tau_{0}\sigma_{0},\quad 
&&\lambda_{3}=-\frac{1}{2}\rho_{0}\tau_{z}\sigma_{0}
\end{eqnarray}
while the axial U(1) one reads
\begin{equation}
\tilde{\Gamma}=\rho_{z}\tau_{z}\sigma_{0}.
\end{equation}
Now one may consider interpolation paths which breaks the doubling-orbital SU(2) symmetry generated by $t_i$ into a doubling-orbital U(1)
Considering a general interpolation between path I and path II by a path which preserves $t_3$, one writes
\begin{eqnarray}
U&=&m\left[\left(\sin\phi t_{3}+\cos\phi\right)i\sin\theta+\cos\theta\right]\\
A&=&-it_{3}A_{o_{3}}-iA,
\end{eqnarray}
where $A_{o_{3}}$ is the gauge field corresponding to the conserved SU(2) component related to the doubled-orbital degree of freedom.

In this basis, we may perform an analysis, similar to what we carried out in the 1+1D case, and introduce the Bardeen form of the Wess-Zumino action in the presence of the $U$ matrix, which captures the system's response due to anomalies. In 3+1 dimensions the result is much more involved, and reads, for vector gauge fields \cite{VPNair}
\begin{eqnarray}
&&\delta S_{WZ}=\nonumber\\
&&(\Gamma_{WZ}\left(A,U\right)-\Gamma_{WZ}\left(A,1\right))/2=\nonumber\\
&&C\int_{M_{4}}Tr\left(dUU^{-1}\right)^{5}\nonumber\\
&&+5C\int_{M_{3}} Tr\left(AdA+dAA+A^{3}\right)\left(dUU^{-1}+U^{-1}dU\right)\nonumber\\
&&-\frac{5C}{2}\int_{M_{3}} Tr\left[AdUU^{-1}AdUU^{-1}-AU^{-1}dUAU^{-1}dU\right]\nonumber\\
&&-5C\int_{M_{3}} TrB\left[\left(dUU^{-1}\right)^{3}+\left(U^{-1}dU\right)^{3}\right]\nonumber\\
&&-5C\int_{M_{3}} Tr\left[dAdUAU^{-1}-dAdU^{-1}AU\right]\nonumber\\
&&-5C\int_{M_{3}} Tr\left[AU^{-1}AU\left(U^{-1}dU\right)^{2}-U^{-1}AUA\left(U^{-1}dU\right)^{2}\right]\nonumber\\
&&+5C\int_{M_{3}} Tr\left(dAA+AdA\right)\left(U^{-1}AU-UAU^{-1}\right)\nonumber\\
&&+5C\int_{M_{3}} Tr\left(AUAU^{-1}AdUU^{-1}+AU^{-1}AUAU^{-1}dU\right.\nonumber\\
&&\left.+\frac{1}{2}UAU^{-1}AUAU^{-1}A\right)
\end{eqnarray}
where $M_3$ and $M_4$ represent the (3+1)D TI manifold and extended (4+1)D manifold for Wess-Zumino-Witten term, respectively. Plugging our specific form for the $U$-matrix one achieves great simplification. The constant $C$ can be determined from the usual chiral anomaly for a doubled TI system via $C Tr\left[\mathbf{1}\right]=\frac{1}{120\pi^{2}}.$ We finally find
\begin{eqnarray}
S_{WZ}&=&\frac{2}{8\pi^{2}}\int \theta_e \left( \vec{x},t\right)
\left(dA_{e}dA_{e}+dA_{o_{3}}dA_{o_{3}}\right)
\nonumber \\
&& +\frac{2}{4\pi^{2}}\int \theta_s \left( \vec{x},t\right)
dA_{e}dA_{o_{3}}
\end{eqnarray} Notice that the $\theta_e$ and $\theta_s$ functions are the same as the ones computed in the 1D scenario in \ref{sub:1dmanes}, except now the full evolution profile $\theta(\vec{x},t)$ may now fluctuate in (3+1)D space, not only in (1+1)D (a note on notation, do not confuse the functions $\theta_e$ and $\theta_s$ with the profile $\theta$ of evolution of the parameter $\alpha$. Compare with equations \eqref{eq:thetae} and \eqref{eq:thetas}). As a consequence, path I binds axion  terms for both electromagnetic and doubling-orbital U(1) gauge fields, while path II binds mixed axion terms between the electromagnetic and doubling-orbital gauge fields.. 

As a check, the integrand of $\theta_e$ in the limiting cases of $\phi=0,\,\pi/2,\,\epsilon$ and $\pi/2-\epsilon$ (for infinitesimal $\epsilon$) can be analytically shown, to first order in $\epsilon$, to reduce exactly the results found from the perturbative diagrammatic approach of the calculation of the electromagnetic part of the action.
More generally, for the evolution to a non-uniform $\theta(\vec{x},t)$  one may use these expressions to study surface effects. For example, if we evolve the system to a profile $\theta(\vec{x},t)$ equal to $\pi$ for $z>0$, however locked to $0$ in for $z<0$, then from Fig. \ref{fig:Ffunc1}, one may write 
\begin{eqnarray}
\theta_e\left(\mathbf{x},t\right)=\pi\Theta\left(z\right)\Theta\left(\pi/4-\phi\right)\\
\theta_s\left(\mathbf{x},t\right)=\pi\Theta\left(z\right)\Theta\left(\phi-\pi/4\right),
\end{eqnarray}
where $\Theta$ is the Heaviside step-function. Plugging into the axion effective actions and integrating by parts will turn the spatial-dependent step functions into Dirac deltas which act to bind remaining part of the effective action to the domain-wall/surface (i.e., $z=0$). The effective description on the surface for evolution on path I of Fig.\ref{fig:Hamilt_space} is electromagnetic and doubled-orbital Chern-Simons terms, while for path II is mixed Chern-Simons between them, as expected from our previous discussions. 

\section{Conclusions \label{sec:conclusion}}
Many decades ago, insulating systems were thought to be uninteresting from the point of view of transport physics. However, Thouless pumps are one example that demonstrate exotic charge transport phenomena in insulators.
By introducing multiple copies of canonical minimal models for Thouless pumps in (1+1)D and (3+1)D, we extended the parameter space for adiabatic Hamiltonian modifications into a larger, multiply-connected space. This is possible because the extra degrees of freedom allow for the addition of different gapping terms to the Hamiltonian which can compete with one another depending on their symmetry properties. Regions of parameter space where the competing terms compensate each other completely define gapless singular regions that must be avoided during pumping processes.

We found that, remarkably, the effective electromagnetic action built in an adiabatic evolution in an open path in such a parameter space depends on the class of path chosen and its corresponding broken symmetry, instead of simply the starting and ending reference Hamiltonians. Full cycle evolutions back to the starting point in parameter space may then define quantum pumps of different quantum numbers and depend on the class of closed loop. In the (1+1)D context with an added spin U(1) symmetry, we found three different pumping processes (i) a charge pump, (ii) a spin pump, and (iii) for a path encircling a singular gapless line, the pumping of a full electron (charge plus spin). From a microscopic point of view, the adiabatic evolutions of the Hamiltonian lead to different, degenerate, ground states introduced by the degrees of freedom from doubling.

The (3+1)D case led to similar results as in (1+1)D, with some reinterpretation of the meaning of the degrees of freedom of the doubled system. In particular, the possibility of realizing the pumps in a mirror-symmetric crystalline topological insulator context implies that the SU(2) (valley) gauge field discussed above may be generated by strain fields. The considerations in this 3D scenario are reminiscent of, and enforce, the concept of Chern-Simons axion pumping described in Ref. \onlinecite{Taherinejad}. We also note that, when viewed from the open path/effective response theory point of view, our Hamiltonian transformations present a realization of the $T$ and $S$ transformations for $Sp(2N,\mathbb{Z})$ actions proposed in Ref. \onlinecite{dimofte} for general (3+1)D $U(1)^N$ Abelian gauge theories (with $N=2$ here). These generalize electric-magnetic duality on the space of conformally invariant boundary conditions for a free $U(1)^N$-flavor Abelian (3+1)D gauge theory. The $Sp(2N,\mathbb{Z})$ has three generators, of "T-,S-, and GL-"types which 
should act at the (3+1)D system boundary. While rotations of type of path-I correspond to T-type transformations, which simply adds to the conserved currents the Hodge dual of the field strength without changing the underlying theory (i.e., it transforms the action by adding the corresponding surface Chern-Simons terms), S-type transformations can be realized by rotations of type-II in our formalism. This adds to the surface theory, as we have seen, mixed Chern-Simons terms, which in fact make the background gauge fields actually dynamical. Searching then for actual physical realizations of such systems as presented here, with surfaces gapped  spontaneously or explicitly with order parameters in the doubling degrees of freedom may be an interesting route to novel topological phenomena.

\section*{Acknowledgements}
We would like to thank M. Stone and V. P. Nair for crucial contributions regarding the Ma\~{n}es-Bardeen actions. PLSL acknowledges support from FAPESP under grant 2009/18336-0
This work was supported in part by the National Science Foundation grant DMR-1455296 (SR) at the University of Illinois, and by Alfred P. Sloan foundation. PG acknowledge support from 
NSF EFRI -1542863. TLH is supported by the ONR YIP Award N00014-15-1-2383.

\appendix
\section{\\Triangle Anomaly Computations \label{app:Appendix-A}}

Let us explicitly present the calculations considered throughout this work
related to the interpolations from Fig. (\ref{fig:Hamilt_space}).
We consider paths $I$ and $III$ in 1D. The calculations for the
other paths, and for the 3D case, follow analogously to these results
summarized in Tables \ref{fig:-Summary1D} and \ref{fig:-Summary3DBdG}.
Calculations are done in Minkowski space with $i\epsilon$ prescription
omitted. This can be taken into account by Wick rotation in the momentum
integrations whenever necessary.

Quite generally, we compute effective actions by computing, in the gradient expansion (over external fields), the partition functions in the presence of a generic electromagnetic gauge field $A,$ and considering small transformations over a smooth and slowly varying path parametrization variable $\alpha$ as
\begin{equation}
S_{eff} [A,\alpha+\delta \alpha] = -i\log \frac{\mathcal{Z}[A,\alpha + \delta \alpha]} {\mathcal{Z}[A,\alpha]}
\end{equation}
which we expand to first order in $\delta \alpha$ considering $\alpha\sim const.$. This is arranged so that we can use small transformations of the Hamiltonian and develop them, step-by-step, into a full profile $\theta(x,t)$,  as discussed in the Introduction. Importantly,  $\theta(x,t)$ may have a strongly fluctuating space-time dependence (like a step function).  For the full profile $\theta$, we then have
\begin{equation}
S_{eff} [A,\theta]=\int_0^{\theta(x,t)} S_{eff} [A,\alpha] \delta \alpha.
\end{equation}

\subsubsection*{Path $I$ }

Consider Hamiltonian (\ref{eq:doubSSH}) with $\boldsymbol{\Delta}=0$,
$m_{5}=m,$ and introduce a path connecting the different insulating
phases as in (\ref{eq:1DChiralPath}), which we repeat here for convenience
for the reader,
\begin{eqnarray}
H & = & \Gamma_{1}\left(p_{x}-eA\right)+\Gamma_{2}m\cos\alpha+\Lambda_{5}m\sin\alpha\nonumber \\
 & = & \Gamma_{1}p_{x}-\Gamma_{1}A+m\Gamma_{2}e^{i\gamma_{5}\alpha}
\end{eqnarray}
with $i\gamma_{5}=\Gamma_{2}\Lambda_{5}=i\Gamma_{1}$.

Now let us introduce a small change in $\alpha\rightarrow\alpha+\delta\alpha$
and unwind the mass phase by rotating the spinors as
\begin{eqnarray}
\psi & = & e^{-i\gamma_{5}\alpha/2}\psi^{'}\nonumber \\
\psi^{\dagger} & = & \psi^{'\dagger}e^{i\gamma_{5}\alpha/2}.
\end{eqnarray}
Equality of the partition functions before and after the fermion rotation leads to the "naive" chiral Ward-Takahashi
identity implying the conservation of the chiral current for vanishing TI mass. We know that this conservation is
actually spoiled due to the chiral anomaly, captured by a non-invariance of the Jacobian. In the present approach, $\alpha$  is considered, overall, a constant and builds no Jacobian. On then just needs to compute the effective action in gradient expansion in $\delta \alpha$ (Goldstone-Wilczek calculation.)

Up to the neglected gradients of $\alpha$, the Hamiltonian thus reads:
\begin{equation}
H=\Gamma_{1}p_{x}-\Gamma_{1}A+\Gamma_{2}m+im\Gamma_{2}\gamma_{5}\delta\alpha.
\end{equation}
The Lagrangian becomes
\begin{eqnarray}
\mathcal{L} & = & \Psi^{\dagger}\left(i\partial_{t}-\Gamma_{1}p_{x}+\Gamma_{1}A-\Gamma_{2}m-im\Gamma_{2}\gamma_{5}\delta\alpha\right)\Psi\nonumber \\
 & = & \bar{\Psi}\left(i\gamma^{\mu}\partial_{\mu}-m+\gamma^{\mu}A_{\mu}-im\gamma_{5}\delta\alpha\right)\Psi\label{eq:Lag1DChiral}
\end{eqnarray}
with $\bar{\Psi}=\Psi^{\dagger}\Gamma_{2}$ and
\begin{eqnarray}
\gamma^{0} & = & \sigma_{0}\tau_{x}\nonumber \\
\gamma^{1} & = & -i\sigma_{0}\tau_{y}\nonumber \\
\gamma_{5} & = & \sigma_{0}\tau_{z}.
\end{eqnarray}
The resolvent reads
\begin{equation}
G_{0}\left(k\right)=\frac{i}{\gamma^{\mu}k_{\mu}-m}=i\frac{\gamma^{\mu}k_{\mu}+m}{k^{2}-m^{2}},
\end{equation}
and the potential is
\[
V=-\gamma^{\mu}A_{\mu}+im\gamma_{5}\delta\alpha.
\]

Terms in the effective gauge action at first order in $\delta \alpha$ give the (topological) contributions to
the average of the chiral current. After integrating out the fermions with (\ref{eq:Lag1DChiral}) in the path integral,
the effective action to second order in the potential $V$ is given by
\begin{equation}
\delta S_{eff}=-i\frac{1}{2}Tr\left(G_{0}VG_{0}V\right).
\end{equation}
Keeping one contribution from the dynamical $\delta\alpha,$ and one from the gauge field $A,$ reduces the
effective action to
\begin{eqnarray}
\delta S_{topo} & = & -iTr\left(G_{0}\left(-{\not}A\right)G_{0}\left(im\right)\gamma_{5}\delta\alpha\right)\nonumber \\
 & = & -m\int_{k}\int_{q}tr\left[G_{0}\left(k\right){\not}A\left(-q\right)G_{0}\left(q+k\right)\gamma_{5}\delta\alpha\left(q\right)\right]\nonumber \\
 & = & -m\int_{q}A_{\mu}\left(-q\right)\delta\phi\left(q\right)\nonumber \\
 &  & \quad
 \times \int_{k}tr\left[G_{0}\left(k\right)\gamma^{\mu}G_{0}\left(q+k\right)\gamma_{5}\right].
\end{eqnarray}
The trace over the spin matrix $\sigma_{0}$ is trivial and gives
a factor of 2. Also, the only non-vanishing trace is the one with
$\tau_{x}\tau_{y}\tau_{z}$, with $tr\left[\gamma^{\mu}\gamma^{\nu}\gamma_{5}\right]=4\epsilon^{\mu\nu}$.
We find
\begin{eqnarray}
 &  & tr\left[G_{0}\left(k\right)\gamma^{\mu}G_{0}\left(q+k\right)\gamma_{5}\right]\nonumber \\
 & = & -tr\left[\frac{\gamma^{\sigma}k_{\sigma}+m}{k^{2}-m^{2}}\gamma^{\mu}\frac{\gamma^{\nu}\left(q+k\right)_{\nu}+m}{\left(q+k\right)^{2}-m^{2}}\gamma_{5}\right]\nonumber \\
 & = & -4m\epsilon^{\mu\nu}\frac{1}{k^{2}-m^{2}}\frac{1}{\left(q+k\right)^{2}-m^{2}}q_{\nu}.
\end{eqnarray}
To the lowest order in $q$ (gradient expansion) we have
\begin{eqnarray}
\delta S_{topo} & = & 4m^{2}\epsilon^{\mu\nu}\int_{q}A_{\mu}\left(-q\right)\delta\alpha\left(q\right)q_{\nu}\int_{k}\frac{1}{\left(k^{2}-m^{2}\right)^{2}}\nonumber \\
 & = & -i4\frac{1}{4\pi}\epsilon^{\mu\nu}\int_{q}A_{\mu}\left(-q\right)\delta\alpha\left(q\right)q_{\nu}\nonumber \\
 & = & -\frac{1}{\pi}\epsilon^{\mu\nu}\int_{x}\delta\alpha\partial_{\nu}A_{\mu},
\end{eqnarray}
in agreement with (\ref{eq:1DpathI}).

\subsubsection*{Path $III$}

So far the results are standard. Now let us consider the tilting of
path $I$ given by path $III$. The way to proceed is as follows.
Hamiltonian (\ref{eq:1DChiralPath}) now acquires a contribution from
$\Delta_{z}\equiv\Delta$,
\begin{eqnarray}
H & = & \Gamma_{1}p_{x}-\Gamma_{1}A+\Gamma_{2}\cos\phi m+\Lambda_{5}\sin\phi m+\Delta\Lambda_{z}\sin\phi\nonumber \\
 & = & \Gamma_{1}p_{x}-\Gamma_{1}A+m\Gamma_{2}e^{i\gamma_{5}\phi}+i\Delta\Gamma_{2}\Gamma_{5}\sin\phi,
\end{eqnarray}
with $\Gamma_{5}\equiv-i\Gamma_{2}\Lambda_{z}=\sigma_{z}\tau_{z}=\sigma_{z}\gamma_{5}$.
Once again introducing small deviations in $\alpha$ leads to
\begin{eqnarray}
H & = & \Gamma_{1}p_{x}+m\Gamma_{2}e^{i\gamma_{5}\alpha}+i\Delta\Gamma_{2}\Gamma_{5}\sin\alpha.\nonumber \\
 &  & -\Gamma_{1}A+m\Gamma_{2}e^{i\gamma_{5}\alpha}i\gamma_{5}\delta\alpha\nonumber\\&+&i\Delta\Gamma_{2}\Gamma_{5}\cos\phi\delta\alpha
\end{eqnarray}
and rotating the spinors, neglecting its spatial and temporal derivatives,
gives
\begin{eqnarray}
H & = & \Gamma_{1}p_{x}+m\Gamma_{2}+i\Delta\Gamma_{2}\Gamma_{5}\sin\alpha e^{-i\gamma_{5}\alpha}\nonumber \\
 &  & -\Gamma_{1}A+m\Gamma_{2}i\gamma_{5}\delta\alpha\nonumber\\&+&i\Delta\Gamma_{2}\Gamma_{5}e^{-i\gamma_{5}\alpha}\cos\phi\delta\alpha.
\end{eqnarray}
The Lagrangian becomes
\begin{eqnarray}
\mathcal{L} & = & \bar{\Psi}\left(i\gamma^{\mu}\partial_{\mu}-m-i\Delta\sin\phi\Gamma_{5}e^{-i\gamma_{5}\alpha}+\gamma^{\mu}A_{\mu}\right)\Psi\nonumber \\
 &  & +\bar{\Psi}\left(-im\gamma_{5}-i\Delta\Gamma_{5}e^{-i\gamma_{5}\alpha}\cos\alpha\right)\delta\alpha\Psi.
\end{eqnarray}
We are going to consider $\Delta\ll m$ and calculate a perturbative
correction to the topological term. The expectation is that this correction
should vanish since this modified path should give the same result.

In computing the approximate Green function we use that, given
a matrix $A$ and a unitary transformation $U$ that diagonalizes
it as $UAU^{\dagger}=D$, we can write
\begin{eqnarray}
1+A & = & U^{\dagger}\left(1+D\right)U\nonumber \\
\Rightarrow\frac{1}{1+A} & = & U^{\dagger}\frac{1}{1+D}U
\end{eqnarray}
and since we can use Taylor expansion to every eigenvalue of
$D$ (or $1+D$),
\begin{eqnarray}
\frac{1}{1+A} & = & U^{\dagger}\frac{1}{1+D}U\nonumber \\
 & \approx & U^{\dagger}\left(1-D\right)U\nonumber \\
 & = & 1-U^{\dagger}DU\nonumber \\
 & = & 1-A.\label{eq:expansion}
\end{eqnarray}

The Green function is, to first order in $\Delta$,
\begin{eqnarray}
G_{0} & = & i\frac{i}{\gamma^{\mu}k_{\mu}-m-i\Delta\Gamma_{5}\sin\phi e^{-i\gamma_{5}\phi}}\nonumber \\
 & = & i\frac{\gamma^{\mu}k_{\mu}+m-i\Delta\Gamma_{5}\sin\phi e^{i\gamma_{5}\phi}}{k^{2}-m^{2}-\Delta^{2}-2m\Delta\Gamma_{5}\gamma_{5}\sin^{2}\phi}.
\end{eqnarray}
Now we notice that $\Gamma_{5}$ and $\gamma_{5}$ commute and may
be diagonalized simultaneously. So we may use (\ref{eq:expansion}),
\begin{eqnarray}
G_{0} & = & i\frac{\gamma^{\mu}k_{\mu}+m-i\Delta\Gamma_{5}\sin\phi e^{i\gamma_{5}\phi}}{k^{2}-m^{2}}\frac{1}{1-\frac{\Delta^{2}+2m\Delta\Gamma_{5}\gamma_{5}\sin^{2}\phi}{k^{2}-m^{2}}}\nonumber \\
 & \approx & i\frac{\gamma^{\mu}k_{\mu}+m-i\Delta\Gamma_{5}\sin\phi e^{i\gamma_{5}\phi}}{k^{2}-m^{2}}\left(1+\frac{2m\Delta\Gamma_{5}\gamma_{5}\sin^{2}\phi}{k^{2}-m^{2}}\right)\nonumber \\
 & \approx & i\frac{\gamma^{\mu}k_{\mu}+m}{k^{2}-m^{2}}\nonumber \\
 &  & -\Delta\sin\phi i\left[\frac{i\Gamma_{5}e^{i\gamma_{5}\phi}}{k^{2}-m^{2}}-2m\sin\phi\frac{\gamma^{\mu}k_{\mu}+m}{\left(k^{2}-m^{2}\right)^{2}}\Gamma_{5}\gamma_{5}\right]\nonumber \\
 & \equiv & G_{00}-\Delta\sin\phi G_{01},
\end{eqnarray}
with
\begin{eqnarray}
G_{00} & = & \frac{\gamma^{\mu}k_{\mu}+m}{k^{2}-m^{2}}\nonumber \\
G_{01} & = & i\left[\frac{i\Gamma_{5}e^{i\gamma_{5}\phi}}{k^{2}-m^{2}}-2m\sin\phi\frac{\gamma^{\mu}k_{\mu}+m}{\left(k^{2}-m^{2}\right)^{2}}\Gamma_{5}\gamma_{5}\right]\nonumber \\
 & = & \Phi_{1}\Gamma_{5}-\Phi_{2}\left(1+2mG_{00}\right)\Gamma_{5}\gamma_{5},
\end{eqnarray}
and
\begin{eqnarray}
\Phi_{1}\left(k\right) & = & -\frac{\cos\phi}{k^{2}-m^{2}}\nonumber \\
\Phi_{2}\left(k\right) & = & \frac{i\sin\phi}{k^{2}-m^{2}}.
\end{eqnarray}
Also, the potential may be written
\begin{equation}
V=V_{A}+V_{\phi}
\end{equation}
with
\begin{eqnarray}
V_{A} & = & \gamma^{\mu}A_{\mu}\nonumber \\
V_{\phi} & = & -im\gamma_{5}\delta\phi-\Delta i\Gamma_{5}e^{-i\gamma_{5}\phi}\cos\phi\delta\phi\nonumber \\
 & \equiv & V_{\phi0}+\Delta V_{\phi1}.
\end{eqnarray}

This allows us to separate the first order in $\Delta$ (over $m$) correction to the action
\begin{eqnarray}
\delta S_{topo} & = & Tr\left(G_{0}V_{A}G_{0}V_{\phi}\right)\nonumber \\
 & = & \delta S_{topo}^{0}+\Delta\delta S_{topo}^{1}
\end{eqnarray}
where
\begin{eqnarray}
\delta S_{topo}^{0} & = & Tr\left(G_{00}V_{A}G_{00}V_{\phi0}\right)\nonumber \\
\delta S_{topo}^{1} & = & Tr\left(G_{00}V_{0}G_{00}V_{\phi1}\right)\nonumber \\
 &  & -Tr\left(G_{00}V_{0}G_{01}V_{0}\right)-Tr\left(G_{01}V_{0}G_{00}V_{0}\right).\nonumber\\
\end{eqnarray}
We have shown that $\delta S_{topo}^{0}$ gives (twice) the expected  theta-term.
Our goal is to see if there is any correction due to $\delta S_{topo}^{1}$.

Realizing that
\begin{eqnarray}
V_{\phi1} & = & \sigma_{z}\left(-i\Delta\gamma_{5}e^{-i\gamma_{5}\phi}\cos\phi\delta\phi\right)\nonumber \\
G_{01} & = & \sigma_{z}\left(\Phi_{1}\gamma_{5}-\Phi_{2}\left(1+2mG_{00}\right)\right)
\end{eqnarray}
it is easy to see that, since the traces over the $\sigma$ and the
$\tau$ matrices decouple, $\delta S_{topo}^{1}$ vanishes (since the
$\sigma_{z}$ traces vanish).

\bibliographystyle{apsrev}
\bibliography{HamiltSpace}

\end{document}